\documentclass[12pt,preprint]{aastex}
\usepackage{lscape}

\usepackage{amsmath}

\newcommand{\lya}{Ly$\alpha$ }
\newcommand{\ha}{H$\alpha$ }

\newcommand{\zz}{$z\sim2$ }
\newcommand{\zzz}{$z\sim3$ }

\newcommand{\zrange}{$1.9<z<3.3$ }

\shorttitle{Detecting $z>2$ Type IIn Supernovae} 
\shortauthors{Cooke} 

\begin{document}

\title{Detecting $z>2$ Type IIn Supernovae}

\author{Jeff Cooke\altaffilmark{1} \altaffilmark{2}} \affil{The Center
for Cosmology and the Department of Physics \& Astronomy, The
University of California, Irvine, Irvine, CA, 92697-4575}
\email{cooke@uci.edu}

\altaffiltext{1}{Gary McCue Postdoctoral Fellow}
\altaffiltext{2}{Based on observations obtained with
MegaPrime/MegaCam, a joint project of CFHT and CEA/DAPNIA, at the
Canada-France-Hawaii Telescope (CFHT) which is operated by the
National Research Council (NRC) of Canada, the Institut National des
Science de l'Univers of the Centre National de la Recherche
Scientifique (CNRS) of France, and the University of Hawaii. This work
is based in part on data products produced at the Canadian Astronomy
Data Centre as part of the Canada-France-Hawaii Telescope Legacy
Survey, a collaborative project of NRC and CNRS.}

%*********************************************************************

\begin{abstract}

Type IIn supernovae (SNe IIn) dominate the brightest supernova events
in observed FUV flux ($\sim$$1200-2000$\AA).  We show that multi-band,
multi-epoch optical surveys complete to m$_{r}=27$ can detect the FUV
emission of $\sim$25 $z>2$ SNe IIn deg$^{-2}$ yr$^{-1}$ rest-frame
($\sim$10 SNe IIn deg$^{-2}$ yr$^{-1}$ observed-frame) to $4\sigma$ 
using a technique monitoring color-selected galaxies.  Moreover, the 
strength and evolution of the bright emission lines observed in 
low-redshift SNe IIn imply that the \lya emission features in 
$\sim$70\% of $z>2$ SNe IIn are above 8m-class telescope spectroscopic 
thresholds for $\sim$2 yr rest-frame.  As a result, existing 
facilities have the capability to both photometrically detect and 
spectroscopically confirm $z>2$ SNe IIn and pave the way for efficient 
searches by future 8m-class survey and 30m-class telescopes.  The 
method presented here uses the sensitivities and wide-field 
capabilities of current optical instruments and exploits (1) the 
efficiency of $z>2$ galaxy color-selection techniques, (2) the 
intrinsic brightness distribution ($\langle M_B\rangle=-19.0\pm0.9$) 
and blue profile of SNe IIn continua, (3) the presence of extremely 
bright, long-lived emission features, and (4) the potential to detect 
blueshifted SNe Ly-$\alpha$ emission shortward of host galaxy 
Ly-$\alpha$ features.

\end{abstract}


%*********************************************************************

\section{Introduction}% Section 1

Type II supernovae (SNe) are explosive, highly luminous events that
result from the core collapse of massive stars after exhausting their
nuclear fuel.  Observations of the sites of Type II SNe have
identified the progenitors of Type II-P, IIn, and SN 1987A-type events
[see \citet{gal07}, references therein, and \citet{li07}].  Because
these SN types are linked to massive stars, they are sensitive probes
of star formation and fundamental tools for the study of galaxy
formation and evolution.  In this paper, we focus on the Type IIn SNe
[SNe IIn; \citet{sch90,f97}] identified by the very bright, narrow
emission lines observed in their spectra.  The emission lines are
believed to be caused in part by the strong interaction of the SN
ejecta with dense circumstellar material\footnote{Spectroscopic
observations reveal that the emission lines are often comprised of
multiple components; a narrow feature (FWHM $\lesssim200$ km s$^{-1}$)
superposed on an intermediate (FWHM $\sim$1000-2000 km s$^{-1}$) or
sometimes broad base [FWHM $\sim$5000-10000 km s$^{-1}$].  A
discussion of the mechanism behind SNe IIn emission-line profiles is
beyond the scope of this paper.  The aim here is the detection and
confirmation of high redshift SNe IIn via their continua and
emission-line flux.}.  SNe IIn exhibit very blue continua
(\S~\ref{detection}), and although the continua fade over a period of
weeks, the emission lines are observed to remain bright for years.

The luminosities of most types of SNe near maximum brightness provide
rare opportunities to study galactic and individual stellar processes
out to very high redshift.  However, the properties and extinction of
most SNe, and capabilities of current facilities, have prevented
detections beyond $z>2$.  For example, although Type Ia SNe are
intrinsically bright, they have high UV extinction that makes $z>2$
detections in deep optical imaging surveys difficult.  In addition,
the deep wide-field IR surveys necessary to find the brighter
rest-frame optical flux of such events are difficult to perform.  Type
II SNe are intrinsically fainter than Type Ia SNe as a whole, but a
subset of the Type IIL and IIP SNe show less UV extinction and are
bright enough to be detected in deep wide-field optical surveys using
current facilities.  However once detected, the short outburst
duration and relative faintness of their continua make a reliable
spectroscopic confirmation prohibitive.  This is true for Type Ia
detections as well.  Therefore the identification and classification
of these SNe at high redshift must use methods such as photometric
data and model template fitting.

We show that SNe IIn are an exception.  The intrinsic brightness and
blue continua of SNe IIn not only enable photometric detection at
$z>2$ in deep optical surveys but their bright long-lived emission
lines provide a means for spectroscopic confirmation.  Because their
properties enable detection out to $z>2$, they have great utility as a
diagnostic tool to probe cosmic scales.

The progenitors of SNe IIn are believed to be luminous blue variable
stars (LBVs) with main sequence masses of $\gtrsim80M_{\odot}$
[\citet{gal07,vd06}].  Detonations of these short-lived, high mass
stars are direct indicators of recent star formation and can be used
to trace the universal star formation rate (SFR) out to high redshift.
However, to do this accurately, the dust extinction of SNe IIn as
compared to the entire SN population needs to be quantified.
Detections at $z\ge2$ would extend the measurements of the universal
supernova rate (SNR) [e.g., \citet{dahlen04,calura06}], help confirm
or refute observed trends with redshift, and improve subsequent
measurements that rely on these values.  As the progenitor mass range
becomes more refined, SNe IIn detections can provide a constraint on
the upper-end of the initial mass function for targeted $z>2$ galaxy
populations with an assumption of little mass evolution with redshift.
In addition, SNe IIn have the potential to provide an independent
statistical discriminator between the SFRs and star formation
histories of $z>2$ galaxies modeled from photometric data and stellar
synthesis codes [e.g., \citet{aes03,pap01}].  Because these are
core-collapse SNe, their density at high redshift will place
constraints on the relic neutrino count
[e.g.,\citet{f04,ando04,louis04}], important for future facilities
such as Hyper-K \citep{nak03} and UNO \citep{jung00}.  Furthermore, a
refinement of the kinetic energy output and fraction of SNe IIn to the
entire SNe population will enable $z>2$ SNe IIn to act as an anchor to
extrapolate the SN contribution to galactic-scale winds observed in
high redshift galaxies [e.g., \citet{a03,c05}] and help estimate mass
loss.

In this paper, we present a photometric technique monitoring
color-selected galaxies that can detect $z>2$ SNe IIn in reasonable
numbers.  Moreover, we analyze the strength and evolution of bright
optical and FUV emission lines observed in low-redshift SNe IIn and
show that the emission lines of a large fraction of $z>2$ SNe IIn are
above the thresholds of 8m-class telescopes for $\sim$2 years
rest-frame.  Consequently, SNe IIn (and for similar reasons Type IIa,
see \S~\ref{contaminants}) are the only SNe that can be
spectroscopically confirmed at $z>2$ using existing facilities.  The
results of deep optical surveys for $z>2$ SNe IIn will carve out a new
area of high redshift research and provide the framework for routine
surveys using next generation facilities.  Deep wide-field imaging
capabilities of future 8m-class survey telescopes and the imaging and
spectroscopic sensitivities of 30m-class telescopes will not only
greatly increase the density of SNe IIn detections but extend
confirmed detections to higher redshifts (\S~\ref{predictions}).  A
description of our photometric technique and tests using data from the
Canada France Hawaii Telescope Legacy Survey (CFHTLS) are presented in
\S~\ref{detection}.  An analysis of the observed SNe IIn spectroscopic
features, \lya behavior, and contaminants to SNe IIn surveys are
discussed in \S~\ref{spec}.  We estimate in \S~\ref{expectations} the
\zrange SNe IIn SNR, the number of detectable SNe IIn for surveys
complete to $m_{r}=27$, and the observational expectations for the
CFHTLS.  We conclude with a brief summary in \S~\ref{summary}.  We
adopt $\Omega_{M}$=0.3, $\Omega_{\Lambda}$=0.7, $H_0$=70 km s$^{-1}$
Mpc$^{-1}$ cosmology.  All magnitudes, unless otherwise noted, are in
the AB system \citep{f96}.  Finally, we work in the rest-frame of the
SNe but note that effect of time-dilation in the observed frame is
important in detection expectations.

%*********************************************************************

\section{Photometric Detection}\label{detection}% Section 2

SNe IIn are, on average, comparable in optical brightness to Type Ia
SNe [$\langle M_{B}\rangle_{Type-Ia} = -19.31\pm0.03$; e.g.,
\citet{astier06}], but because they have a broader distribution, they
comprise some of the brightest SN events known.  \citet{r02}
determined an extinction corrected value of $\langle
M_B\rangle_{Type-IIn}$ = $-18.97$ (using our cosmology) with a
dispersion of $\sigma_{obs}=0.92$ for eight SNe IIn that had maximum
brightness information in their sample of 46.  In addition, four of
the remaining 38 events with maximum brightness lower limits were
brighter than $M_B$ = $-20$.  The observed uncorrected fit for the
\citet{r02} sample is $\langle M_B\rangle_{Type-IIn}$ =
$-18.60\pm0.31, \sigma=0.92$ and is the value we adopt in this paper
for our calculations.

The FUV photometry and spectroscopic continua acquired to date
indicate that SNe IIn suffer only moderate extinction.  For example,
the Swift observation of SN 2006bv \citep{i06} reports B =
19.0$\pm0.2$ and UVM2 $[166-268 nm]=20.1\pm0.3$ three days after
discovery.  The Swift observation of SN 2007bb \citep{i07} reports B =
17.9 and UVM2 $[166-268 nm] = 19.5$ two weeks after discovery.  In
addition, the FUV spectrum of SN 1998S \citep{f05} shows estimated
magnitudes of $m_{4300}\sim12.2$ and $m_{1700}\sim13.5$ at 14 days
after outburst.  SNe IIn SEDs are modeled as cooling blackbody spectra
with an additional observed UV excess \citep{leonard00,chugai01,f01}.
The observations that SNe IIn are $\sim$1 mag fainter at maximum light
at $\sim$1700\AA~as compared to their B mag, and rather rapid fade in
the FUV $3-4$ weeks afterward, are consistent with this model.  As a
result, {\it SNe IIn are the brightest of any SN type in the FUV}.
This is important for $z>2$ detection via the method presented in this
paper because the rest-frame FUV is redshifted to the optical
(rest-frame flux at $\sim$1700\AA~corresponds to the observed $r$-band
at $z\sim2.6$).  From the above discussion, the observed uncorrected
$\langle M_B\rangle\sim-18.6$ of SNe IIn would correspond to $\langle
M_{1700}\rangle\sim-17.5$ at maximum brightness.  This translates to
$28.0\lesssim m_{r}\lesssim29.5$ over the redshift path $1.9<z<3.3$.
Therefore, detections of SNe IIn in surveys within the flux
uncertainties of complete to m$_{r}=27$ are possible, however, any
significant detection will need to be pulled from the bright end of
the SNe IIn magnitude distribution.

SNe IIn are estimated to comprise $\sim2-5$\% of the total Type II
population \citep{cap97,dahlen04}.  Therefore a large number of
galaxies must be monitored to witness a relatively small number of
events.  Deep wide-field optical surveys can target large volumes at
high redshift and detect the FUV emission of a large number of high
redshift star-forming galaxies.  The widespread Lyman break galaxies
(LBGs) at $z>2$ \citep{s96,s03,s04} comprise a well-studied,
well-controlled high redshift population
\citep{a03,pet01,aes01,aes03,erb04,c05,c06} with many identifying FUV
features shifted to optical wavelengths.

We present a method to detect SNe IIn in $z>2$ LBGs using
straightforward, albeit deep, existing and future multi-epoch optical
surveys (e.g., CFHTLS, Keck Deep Fields, Subaru Deep Field, GOODS, and
HST programs).  LBGs would be initially identified in stacked or deep
epoch images and monitored over multiple epochs to create LBG light
curves.  Transient flux as a result of a SNe IIn could be detected to
a chosen limit of the photometric uncertainty.  Searching for flux
variations in images with a large number of $z>2$ LBGs can take
advantage of relative photometry and avoid the introduction of noise
that appears in the conventional method of image subtraction.  In
addition, this approach filters SNe detections to the targeted
redshifts of the color-selection.  In this paper, we focus on
$1.9<z<3.3$ largely because of (1) the deep wide-field optical imaging
and spectroscopic sensitivities of existing facilities, (2) the
ability to select large number of faint high redshift galaxies via
successfully demonstrated optical photometric color-selection
techniques at \zz \citep{s04} and \zzz \citep{s96,s03,c05}, and (3)
the inherent properties of SNe IIn, but remark that the principles of
this method can be applied to all redshifts.

LBG color-selection techniques require deep multi-band images that can
also be used to photometrically select rest-frame FUV of $z>2$ SNe
IIn.  Although SNe IIn can be modeled as blackbodies as described
above, this is not straightforward largely because few FUV light
curves exist.  In addition, SNe IIn show some diversity in the decay
time of the more abundant optical light curves.  A few SNe IIn exhibit
delayed decline (e.g., SN 1995N) whereas the majority follow profiles
that (1) show relatively slow brightening, (2) remain consistent to
within one mag of their maximum brightness $\sim$50d after outburst,
and (3) decline $\sim$3 mag from their maximum brightness over the
first 100d [e.g., see \citet{li02}].  We note that the mechanism
behind the FUV emission and UV excess mentioned above is not
completely understood and it is unclear at this point if the diversity
observed in the optical light curve decay rates extends similarly to
the FUV.  More FUV data is needed to properly assess this.  In
addition, environment plays a role in their luminosity because the SN
ejecta interact with circumstellar material and create bright emission
lines.  These lines remain bright for years and make a non-negligible
contribution to the overall luminosity in both the optical and FUV.
Even though only coarse FUV color-selection criteria for $z>2$ SNe IIn
have been established at this point, we show in \S\ref{sims} that this
is sufficient because of their mag distribution and blue continua
(\S~\ref{sims}).  We address other transient event detections in
\S~\ref{contaminants}.

This method to detect the FUV continuum flux of SNe IIn in $z>2$ LBGs
uses broad-band filters.  However, the bright emission lines of SNe
IIn make a time-dependent contribution to the FUV flux.  Typical
broad-band filters have a bandpass of $\sim$$700-1500$\AA~that
correspond to $\sim$$150-350$\AA~rest-frame for $1.9<z<3.3$ objects.
Because SNe narrow emission lines are $\sim$20\AA, an alternate or
complementary approach using existing or future facilities would be to
incorporate wide-field multi-epoch narrow- or intermediate-band
filters tuned to the redshifted Ly-$\alpha$ (and Mg\textsc{ii}
$\lambda\lambda 2795.5, 2802.7$, see \S~\ref{contaminants}) emission
features with imaging in an off-band for selection control.

\subsection{Simulations}\label{sims}% Section 2.1

\begin{figure}
\begin{center}
\scalebox{0.6}[0.6]{\rotatebox{90}{\includegraphics{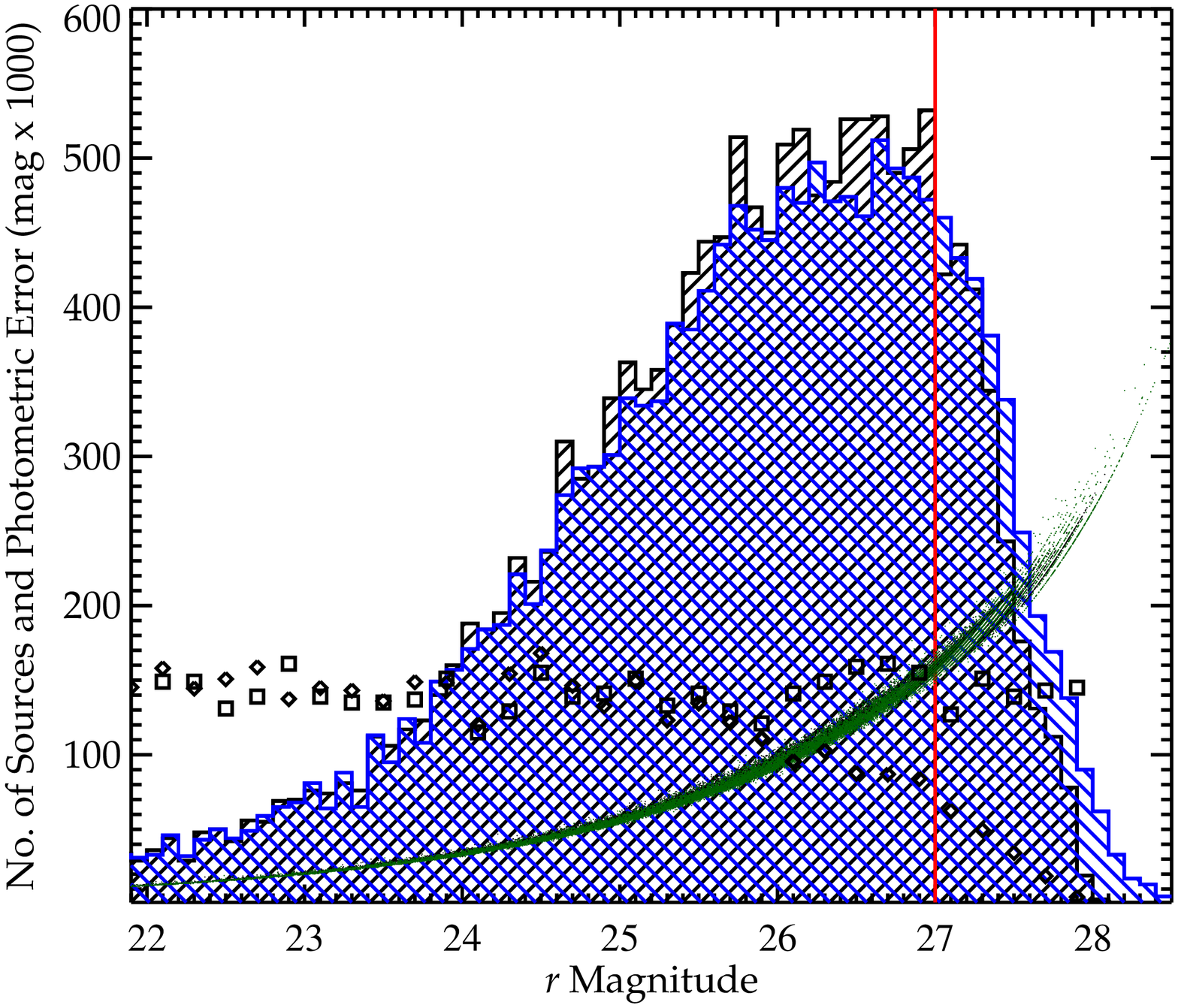}}}
%\scalebox{0.6}[0.6]{\rotatebox{90}{\includegraphics{f1_BW.ps}}}
\caption
{Deep imaging {\it r'} magnitude histogram illustrating the
  completeness of the observational data used for the photometric
  detection simulations. The number of sources per 0.1 mag bin for the
  $\sim17000$ detections in each of two epochs {\it [forward hatch
  (blue) and back hatch (black)]} of the data are shown.  The data are
  complete to {\it r} = 27.0 {\it [vertical {\it (red)} line]} for
  both epochs analyzed.  Here we define completeness as the detection
  of $>$50\% of the simulated sources inserted into the images.  The
  number of simulated sources per 0.2 mag bin is overlaid as squares
  and the number of detected simulated galaxy sources as diamonds
  ($\times 2$).  The discrepancy in the number of sources per bin
  includes the effect of binning caused by difference in magnitude of
  inserted and extracted sources and the ability of the source
  extraction routine to clearly distinguish inserted sources blended
  with existing objects. Overlaid are the photometric errors {\it
  [points (green/black)]} corresponding to their respective {\it r'}
  mag ($\times 1000$).  Sources brighter than {\it r'} $\sim27$ have
  uncertainties $<0.2$ mag.}
\label{fig:rlim}
\end{center}
\end{figure}

The intrinsic brightness, blue SEDs, and bright emission lines of SNe
IIn observed at low to intermediate redshift argue for the feasibility
of their photometric detection at high redshift.  To test this, we
used two $\sim12'\times12'$ deep $r'$-band images taken over two
epochs kindly obtained from the CFHTLS Deep Synoptic Survey
\citep{sul07}.  These deep images are the combined stacked images
taken over several epochs during $5-6$ month observing
seasons\footnote{For more information regarding the CFHT Legacy Survey
and image criteria, see: http://www.cfht.hawaii.edu/Science/CFHLS/}.
We found the seeing to be $\sim0.7$ arcsec FWHM across the images and
a source completeness to m$_{r'}=27.0$, where we define the
completeness as the ability to reliably extract $>50$\% of the
simulated point sources inserted into the images.  This is consistent
with the number density of real sources detected per magnitude
(Figure~\ref{fig:rlim}).

Although photometric detections of SNe IIn and other SNe at
intermediate to high redshifts ($0.6<z<1.7$) will exist in deep
optical survey images, the intrinsic FUV brightness of SNe IIn make
them the dominant SNe detectable at \zrange.  Using SNe templates
modeled from data and available online\footnote{
http://supernova.lbl.gov/$\sim$nugent/nugent\_templates.html}, we
calculate that $\sim$93\% SNe IIn, $\sim$7\% Type IIL, and a
negligible number of Type IIP comprise the detectable SNe at $z>2$ for
optical surveys complete to a magnitude limit of m$_{r}=27$.  To
select $z>2$ SNe, we duplicate the observational process of detecting
SNe IIn in LBGs in the simulations.

LBGs have a half-light radii of $\lesssim0.3$ arcsec at $z>2$ [e.g.,
\citep{s96,g01}] and appear essentially as point sources in
ground-based images [although some appear extended \citep{erb04}].  We
created simulated LBGs and SNe using the PSF of stars convolved with
the seeing in the images using a density similar to the expected
m$_{r'}<27$ LBG density at $1.9<z<3.3$.  In an attempt to mimic actual
observations, we inserted the simulated LBGs randomly, with no attempt
to avoid bright objects or edges.  Although LBGs are known to exhibit
clustering, simulating galaxies with random angular positions is a
reasonable approximation considering the large redshift path probed.

\begin{figure}
\begin{center}
\scalebox{0.35}[0.35]{\rotatebox{90}{\includegraphics{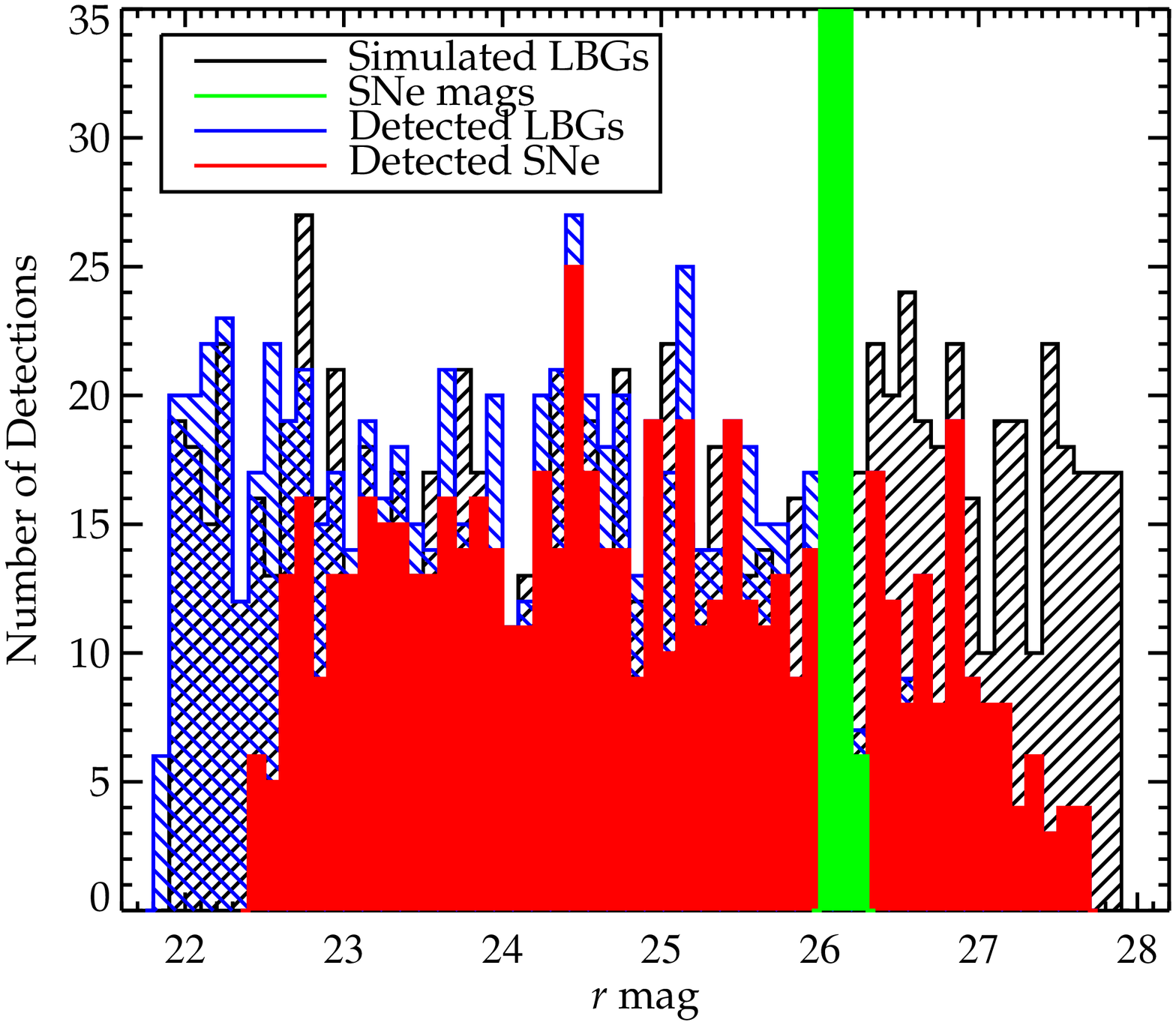}}}
\scalebox{0.35}[0.35]{\rotatebox{90}{\includegraphics{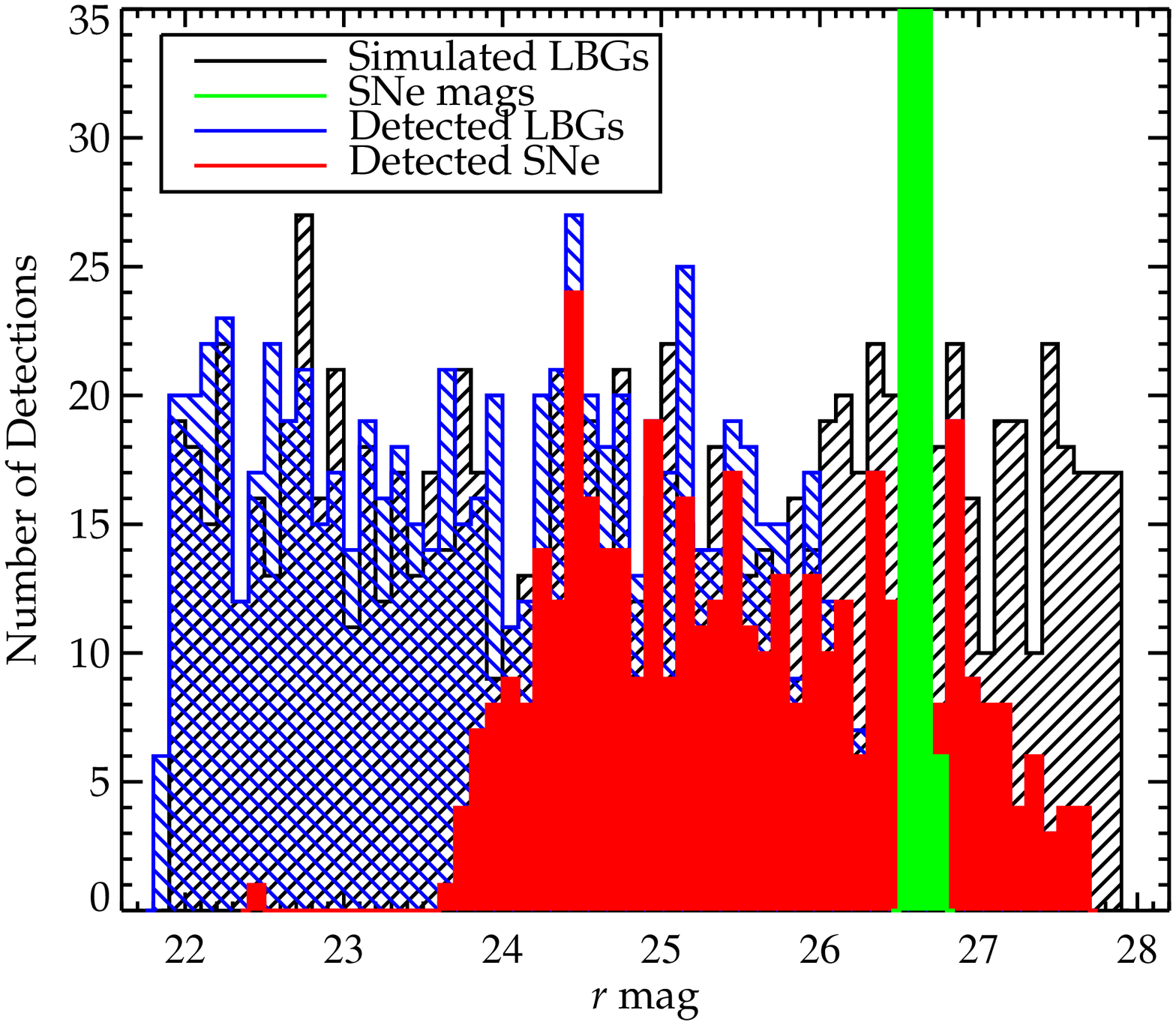}}}
\scalebox{0.35}[0.35]{\rotatebox{90}{\includegraphics{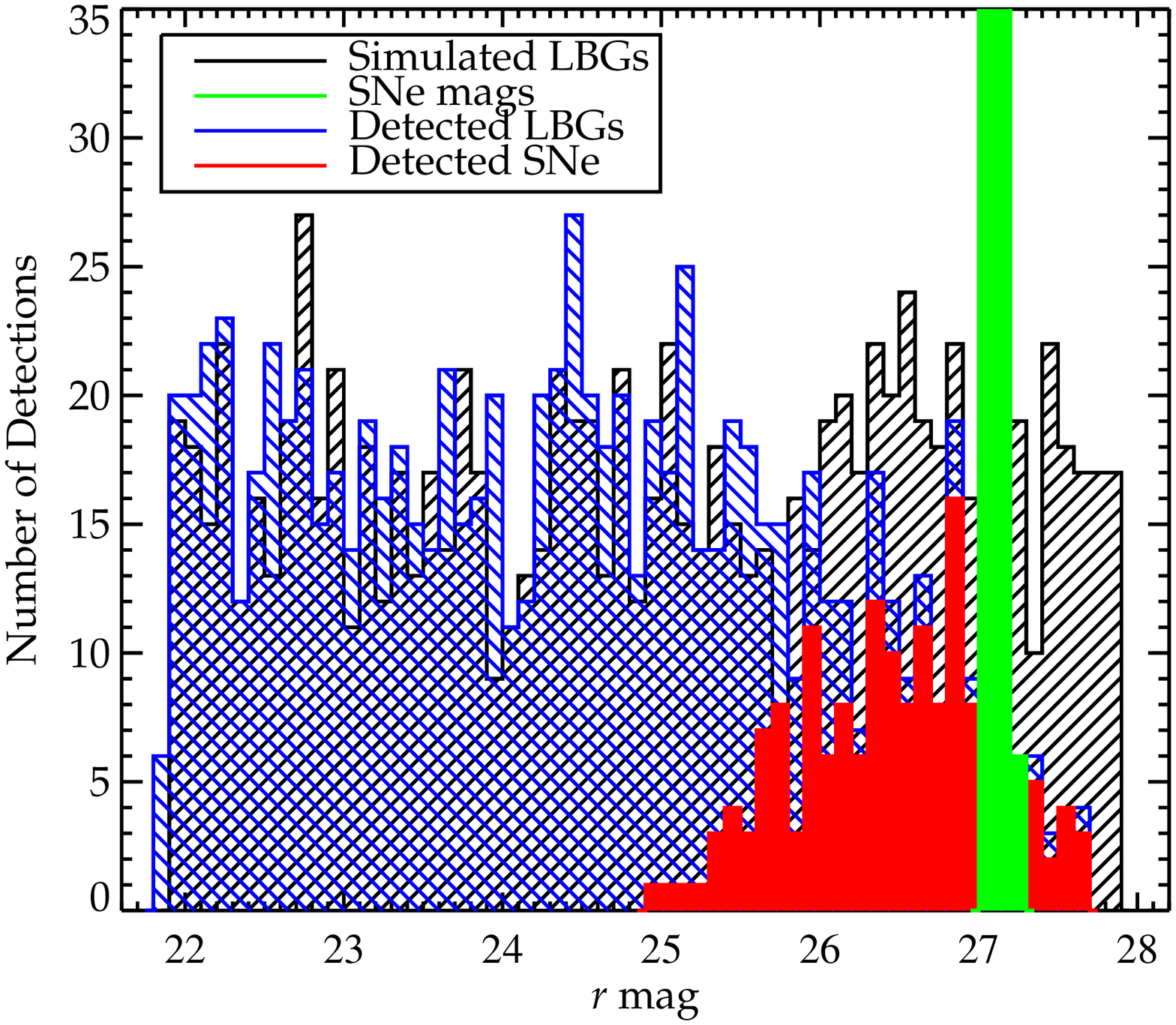}}}
\scalebox{0.35}[0.35]{\rotatebox{90}{\includegraphics{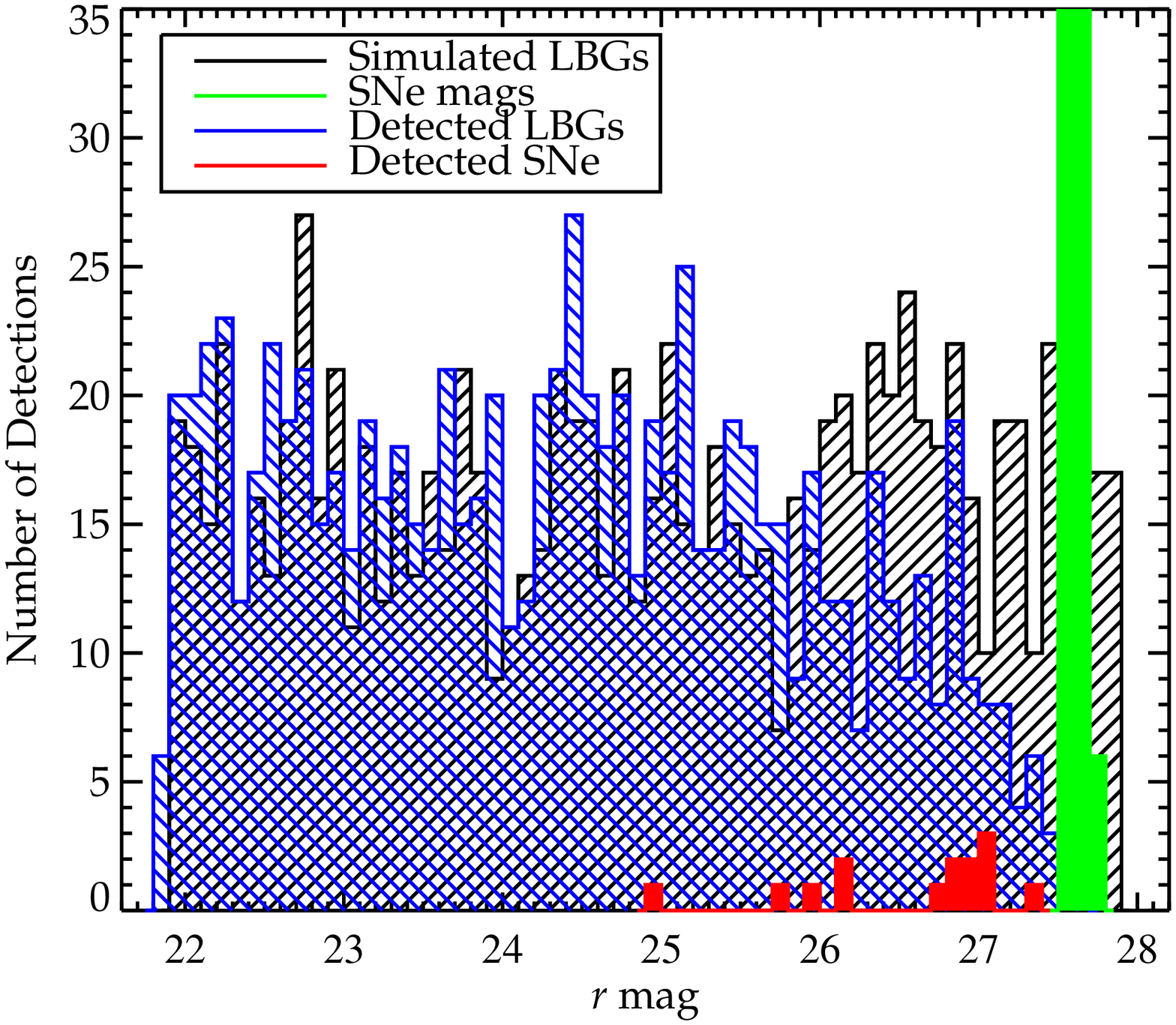}}}
%\scalebox{0.35}[0.35]{\rotatebox{90}{\includegraphics{f2a_BW.ps}}}
%\scalebox{0.35}[0.35]{\rotatebox{90}{\includegraphics{f2b_BW.ps}}}
%\scalebox{0.35}[0.35]{\rotatebox{90}{\includegraphics{f2c_BW.ps}}}
%\scalebox{0.35}[0.35]{\rotatebox{90}{\includegraphics{f2d_BW.ps}}}
\caption
{Detection efficiency histograms for the deep images. The forward
  hatch {\it (black)} histogram indicates the number of simulated LBGs
  per 0.1 mag bin and the back hatch {\it (blue)} histogram shows the
  number of LBGs extracted per mag bin as described in the text. The
  solid light gray {\it (green)} histogram represents the magnitude
  distribution of the simulated SNe for that particular run and the
  solid dark gray {\it (red)} histogram indicates the number of LBGs
  per mag bin with successful SN detections at $\ge4\sigma$ flux
  uncertainty.  For example, {\it (upper left)} SNe with
  m$_{r'}=26.0\pm0.1$ {\it ($-19.8\gtrsim M_{FUV}\gtrsim-21.3$ at
  $1.9<z<3.3$)} are detected in 73\% of monitored $22<m_{r'}<28$ LBGs
  whereas {\it (lower right)} SNe with m$_{r'}=27.5\pm0.1$ {\it
  ($-18.3\gtrsim M_{FUV}\gtrsim-19.8$)} are detected in 2\% of the
  LBGs.}
\label{fig:eff}
\end{center}
\end{figure}

We were able to extract $\gtrsim$90\% of LBGs with $22.0<r'<27.0$
using {\it SExtractor} \citep{ba96} with a detection threshold value
of $1.3\sigma$ and minimum detection area of 10 pixels (0.187 arcsec
pixel$^{-1}$).  Overlaid in Figure~\ref{fig:rlim} are the number of
simulated LBGs and the number of detected LBGs as a function of
magnitude.  These values were optimized in the usual manner by
minimizing spurious detections in negative images.  As expected, the
bulk of the objects that we were unable to extract were those with
$r'>27.0$, with detections falling to near zero at $r'=28.0$, and
$\sim$2\% of all simulated sources falling on bright stars, edges, or
otherwise contaminated regions.  To estimate systematics and determine
LBG detection as a function of magnitude, we used {\it SExtractor} in
dual mode to extract the same LBGs in a second epoch image using the
pixels defined by the segmentation map of the first epoch.  This
duplicates the process by which LBGs are first color-selected for a
given survey at one epoch (or in a combined image) and then monitored
over the remaining (or separate) epochs.  Figure~\ref{fig:eff}
illustrates this analysis for one of the simulations.  In all four
panels of the figure, the number of simulated LBGs are indicated by
the forward hatch (black) histogram and the number of extracted LBGs
are indicated by the back hatch (blue) histogram.  It can be seen that
the detections drop to $\sim50$\% at m$_r'=27$.

To determine the sensitivity of the images for detecting SNe in
$1.9<z<3.3$ LBGs, we overlaid simulated SNe at the locations of the
LBGs in one of the epochs.  We recorded the fraction of SN detections
for a series of SN magnitudes incremented in steps of 0.25 mag with a
distribution of $\pm0.1$ mag.  Figure~\ref{fig:eff} displays the
results of four of these iterations.  In this figure, the dark (red)
solid histogram indicates the number of SNe detections and the light
(green) solid histogram indicates the SNe mag range of the particular
iteration.  The LBG photometry in one epoch were compared to the LBG
photometry in the second epoch that included the simulated SNe.  We
chose to define a successful SNe detection as an LBG that displayed a
$\ge4\sigma$ flux enhancement for that LBG.  The efficiency was
initially determined by placing the simulated SNe at the centroids of
the LBGs.  To test the efficiency for SNe having offsets from host
galaxy centers, the SNe were inserted with radial offsets in
increments of 0.1 arcsec from the LBG centroids.  We found negligible
differences in the detection results up to a radius of 0.5 arcsec,
corresponding to $r\sim15-25 h^{-1}$kpc at $z=1.9-3.3$.  Beyond that,
the overlap of the defined SN pixels with the defined LBG pixels was
reduced to the extent that each SN became better detected as a
transient event separate from the LBG and would need to be detected by
image subtraction.  In practice, a photometric search for transients
as a function of radius about each color-selected LBG centroid could
be employed to select potential $z>2$ SNe from the many transients in
deep wide-fields.  Finally, from the number of successful SN
extractions as a function of LBG and SN magnitudes, we determined the
SNe detection efficiency function for this photometric method.  We use
this function as one of the parameters in estimating observational
expectations in \S\ref{expectations}.  All SNe detections determined
in this manner can be checked by the conventional method of image
subtraction that can also help ascertain the SN centroid offset from
the host galaxy.

\section{Spectroscopic Confirmation}\label{spec}% Section 3

SNe IIn show promise for $z>2$ {\it spectroscopic confirmation}
because their optical and FUV emission lines grow extremely bright and
remain bright for years.  At $1.9<z<3.3$, the primary SNe IIn emission
line in deep optical spectroscopic observations is Ly-$\alpha$.
Currently only a few low redshift FUV spectra of SNe IIn exist.
Therefore we first study the behavior of the larger dataset of
H-$\alpha$ emission lines in the literature and then focus on the
smaller FUV dataset to construct line ratio and Ly-$\alpha$
emission-line strength expectations.

\begin{figure}
\begin{center}
\scalebox{0.6}[0.6]{\rotatebox{90}{\includegraphics{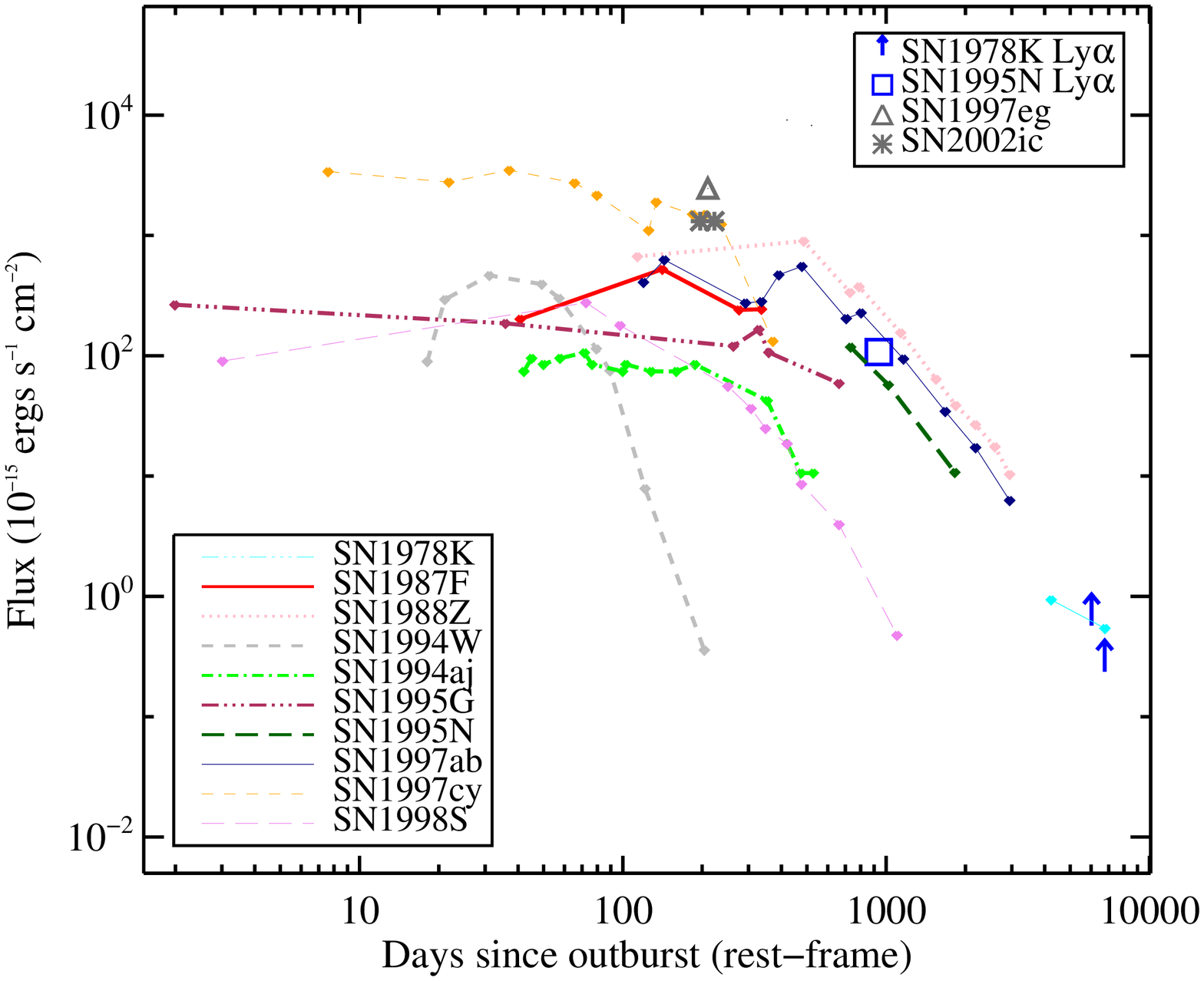}}}
%\scalebox{0.6}[0.6]{\rotatebox{90}{\includegraphics{f3_BW.ps}}}
\caption
{Observed evolution of emission line flux.  Both the flux measurements
  and days after outburst are corrected to the arbitrary epoch of
  $z=0.01$.  Evolution of the H-$\alpha$ emission lines of ten SNe IIn
  [SN 1978K \citep{sch99,ch95}; SN 1987F \citep{f89}; SN 1988Z
  \citep{t93,a99}; SN 1994W \citep{ch04}; SN 1994aj \citep{b98}; SN
  1995G \citep{p02}; SN 1995N \citep{f02}; SN 1997ab \citep{h97}; SN
  1997cy \citep{asiago00}; and SN 1998S \citep{f01}] are indicated by
  diamonds and notated curves.  H-$\alpha$ emission line values for
  two Type IIa SNe [SN 1997eg \citep{sal02} and SN 2002ic \citep{d04}]
  are shown by the {\it (green)} asterisks and triangle,
  respectively. Values for the Ly-$\alpha$ emission lines of two SNe
  [SN 1995N \citep{f02} and SN 1978K \citep{sch99,ch95}] are shown as
  {\it (blue)} squares. }
\label{fig:z01sn_flux}
\end{center}
\end{figure}

The brightness evolution (or single measurement) of the H-$\alpha$
emission line flux of ten SNe IIn (and two Type IIa SNe) is shown in
Figure~\ref{fig:z01sn_flux}.  In this figure, all fluxes and days
after outburst are corrected to the arbitrary epoch of $z=0.01$.  The
SNe used here are assumed to be representative of typical SNe IIn.
They have observed absolute B magnitudes near maximum brightness that
are consistent with the distribution adopted in this paper [SN1978K
$<-12.2$; SN1987F $=-18.7$; SN1988Z $=-18.6$; SN1994W $=-18.6$;
SN1994aj $=-17.7$; SN1995G $=-18.6$; SN1995N $=-14.6$; SN1997ab
$=-19.0$; SN1997cy $=-19.9$; SN1998S $=-18.9$ (Type IIa: SN1997eg
$=-18.0$; SN2002ic $=-19.9$)]. It can be seen in
Figure~\ref{fig:z01sn_flux} that, except for the rapid decline
observed in SN1994W (which also has a reported rapid decline in its
optical continuum), there is a general form to the emission line
evolution. The H-$\alpha$ emission line flux appears (1) to remain
bright or increase in brightness for the first $\sim200$d, (2)
``knee'' or peak in brightness, and (3) exhibit a slow consistent
decline.  This trend is seen for each SNe IIn but at different
observed strengths.

\begin{figure}
\begin{center}
\scalebox{0.4}[0.4]{\rotatebox{90}{\includegraphics{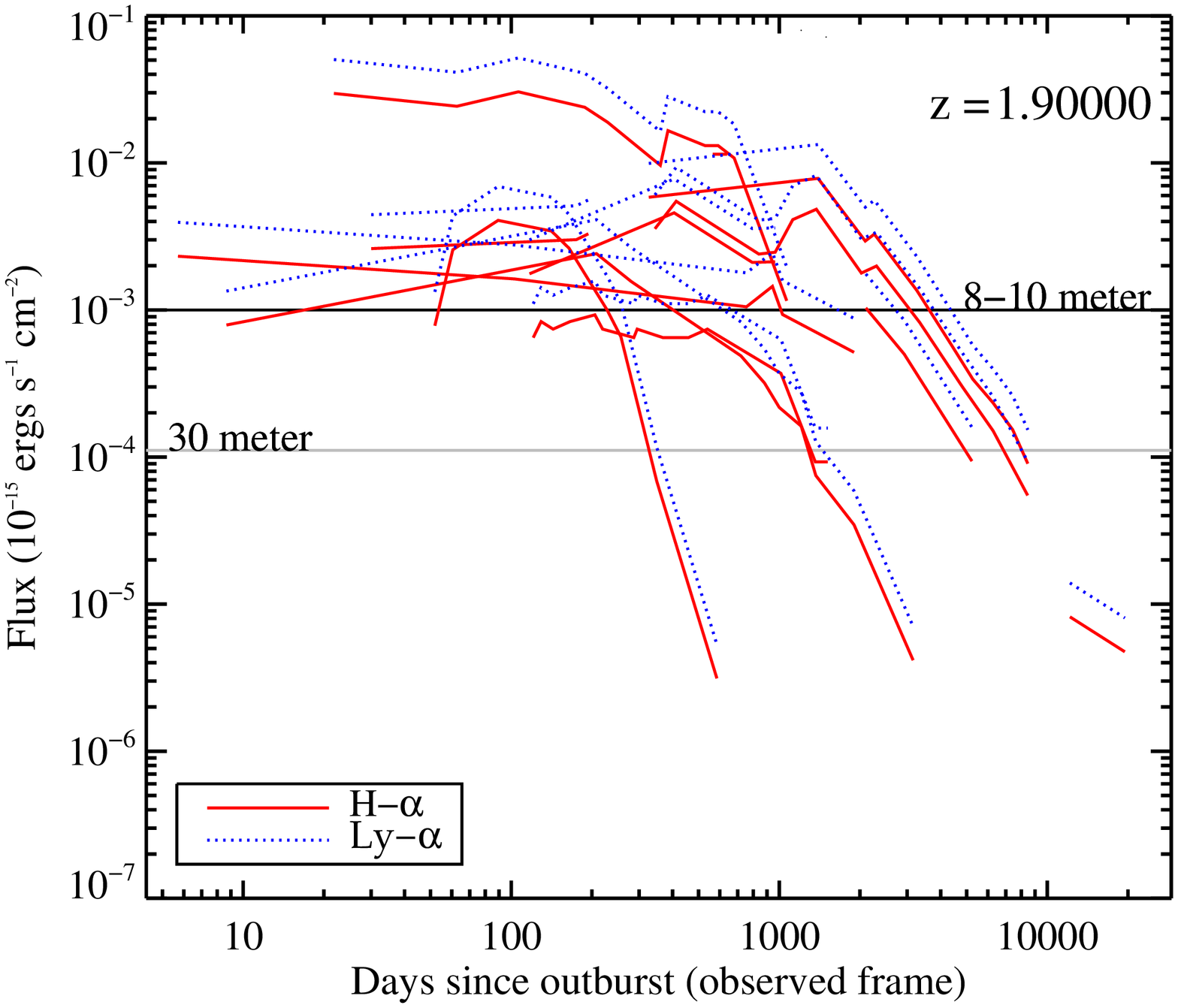}}}
%\scalebox{0.45}[0.45]{\rotatebox{90}{\includegraphics{f4a_BW.ps}}}
\scalebox{0.4}[0.4]{\rotatebox{90}{\includegraphics{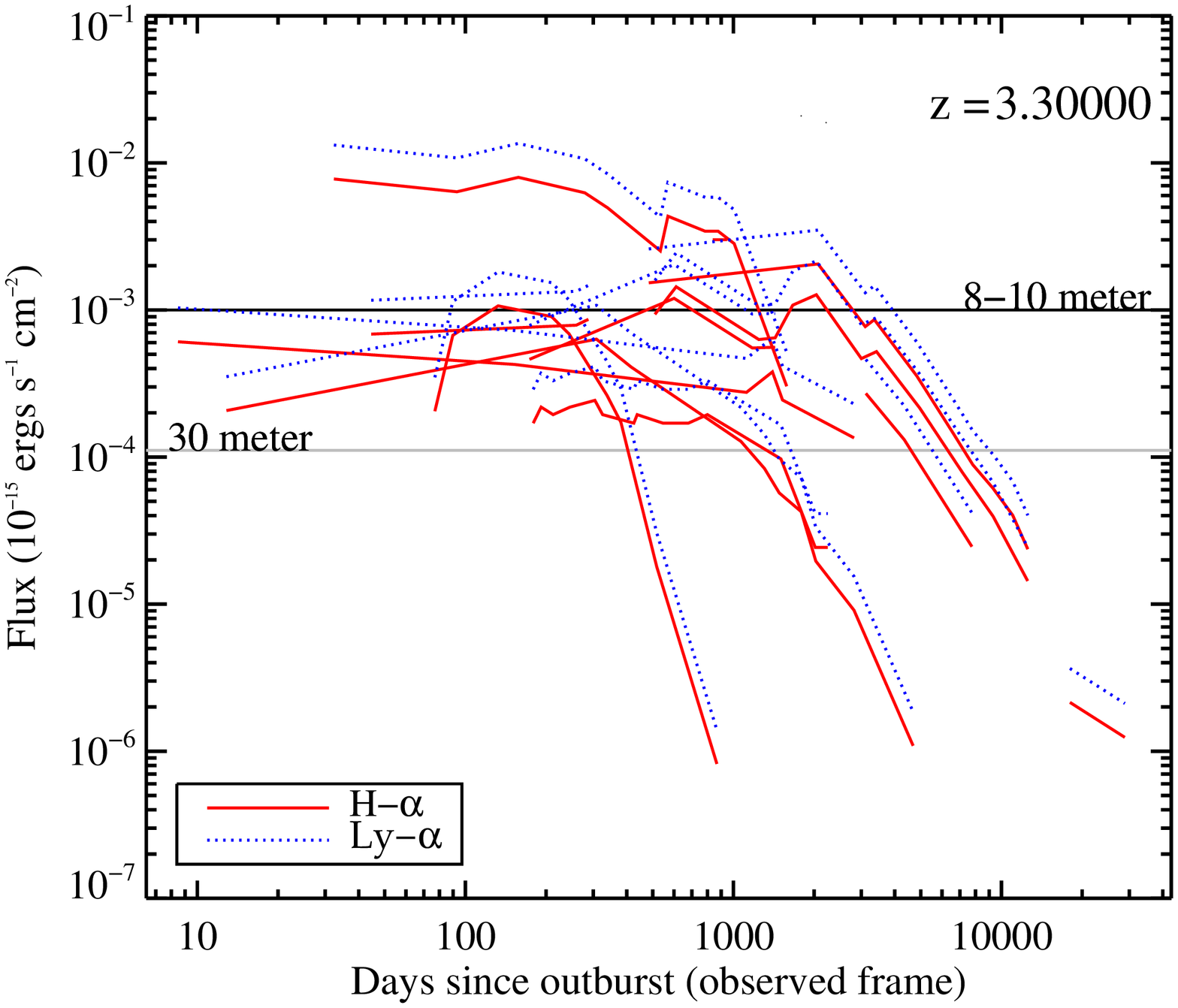}}}
%\scalebox{0.45}[0.45]{\rotatebox{90}{\includegraphics{f4b_BW.ps}}}
\caption
{Estimate of the spectroscopic detectability of $z>2$ Type IIn SNe
  emission lines.  These plots are the similar to
  Figure~\ref{fig:z01sn_flux} but with the H-$\alpha$ emission line
  flux evolution values shown as solid {\it (red)} curves redshifted
  to $z=1.9$ {\it (upper panel)} and $3.3$ {\it (lower panel)}.  The
  dotted {\it (blue)} curves estimate the Ly-$\alpha$ emission line
  flux evolution for the ten SNe using the Ly-$\alpha$/H-$\alpha$ line
  ratio found in \citet{f02} and assumed constant.  The sensitivity of
  existing instruments on 8m-class telescopes with long {\it ($\sim$4
  hr.)}  integrations is indicated by the upper horizontal solid line
  and an estimate of the sensitivity of a future 30m-class telescope
  is indicated by the lower horizontal solid line as labeled.  Note
  the long duration in which the emission lines of $z>2$ SNe IIn are
  detectable in the observer frame.}
\label{fig:obj1250_z3}
\end{center}
\end{figure}

The peak H-$\alpha$ emission line fluxes ($f_{H_\alpha}$) in these
low- to intermediate-redshift SNe IIn would be between
$10^{-17}\lesssim f_{H_\alpha} \lesssim 10^{-18}$ erg s$^{-1}$
cm$^{-2}$ if observed at $1.9<z<3.3$.  Previous long exposure {\it
($\sim$2-6 hr.)} observations [e.g., \citet{martin04,c05} and
\citet{v07}] using LRIS \citep{oke95} on Keck and FORS2 \citep{ap98}
on the VLT have demonstrated the ability to detect narrow emission
line fluxes as faint as $\sim$1$\times 10^{-18}$ erg s$^{-1}$
cm$^{-2}$ to $\sim 3 \sigma$\footnote{This includes observations using
LRIS with the new, higher sensitivity, blue CCDs.}.  The \lya features
of SNe IIn are expected to be equivalent to or brighter than the \ha
features.  The Ly-$\alpha$/\ha line ratio of SN1995N is observed to be
$\sim1.7$ \citep{f02} and $\sim$1 for SN1978K \citep{sch99}.  However,
the latter is more difficult to assess relative to the discussion here
because of the pointing errors reported in that work and its late
stage of evolution.  Figure~\ref{fig:obj1250_z3} shows the values for
the SNe plotted in Figure~\ref{fig:z01sn_flux} if observed at $z=1.9$
and $z=3.3$.  The \lya evolution in Figure~\ref{fig:obj1250_z3} is
plotted to follow the \ha evolution with a constant Ly-$\alpha$/\ha
ratio of 1.7.  In this manner, the evolution of the \ha and \lya line
strengths shown here bracket the observed values for \zrange SNe IIn
\lya strengths.  We have a {\it Hubble Space Telescope} program
underway to study a sample of low to intermediate redshift SNe IIn to
improve the FUV continuum and Ly-$\alpha$ emission line measurements
and the Ly-$\alpha$/H-$\alpha$ ratio and evolution.

It is important to note that at $1.9<z<3.3$, Ly-$\alpha$ features are
redshifted between $3500-5200$\AA~and fall over a wavelength range
where there are no bright night sky emission lines to conflict with
detections.  In addition, the Ly-$\alpha$ emission lines of
$1.9<z<3.3$ SNe IIn are estimated to remain detectable for $\sim$2
years in the rest-frame, or $\gtrsim6$ years in the observer's frame.
Integrating the fraction of photometrically detectable SNe over
redshift path $1.9<z<3.3$, we find the bulk of detectable SNe IIn near
$z\sim2$.  Therefore, an extrapolation of the \ha evolution using a
Ly-$\alpha$/\ha emission line flux ratio of $\sim$1 results in
$\gtrsim$85\% of all \zrange SNe IIn similar to this sample would have
detectable \lya emission during the brightest epoch of their evolution
using current facilities.  In the case for 30-m telescopes, all SNe
IIn similar to this sample would be detectable, and for a longer
duration.

\subsection{Ly-$\alpha$ emission}\label{lya}% Section 3.1

Beyond their FUV luminosity and emission-line strength, another
advantageous aspect toward the detection of $z.2$ SNe IIn is that
their emission lines are observed to be blueshifted with respect to
the host galaxy by $\sim$$1500-3500$ km s$^{-1}$ [e.g.,
\citet{f02,f05,t93,f97,leonard00}] and can aid in feature separation.
If this weren't the case, the intrinsic Ly-$\alpha$ emission and/or
absorption observed in $z>2$ LBG spectra could present a serious
obstacle.  Although the H-alpha line profiles are sometimes observed
to decompose into multiple components, the Ly-alpha feature appears as
a single component and is observed to remain blueshifted through day
494 for SN 1998S \citep{f05} and day 943 for SN 1995N \citep{f02}.
The \lya emission features of the host LBGs are redshifted $\sim$600
km s$^{-1}$and the \lya absorption features are blueshifted $\sim$150
km s$^{-1}$ on average with respect to the systemic redshift
\citep{a03,aes03,c05}.  The systemic offset of LBG Ly-$\alpha$
features has been determined using nebular emission line observations
\citep{pet01} and is attributed to galactic-scale outflows driven by
stellar and SNe winds [also see \citet{m07,tap07}].

\begin{figure}
\begin{center}
\scalebox{0.5}[0.37]{\rotatebox{90}{\includegraphics{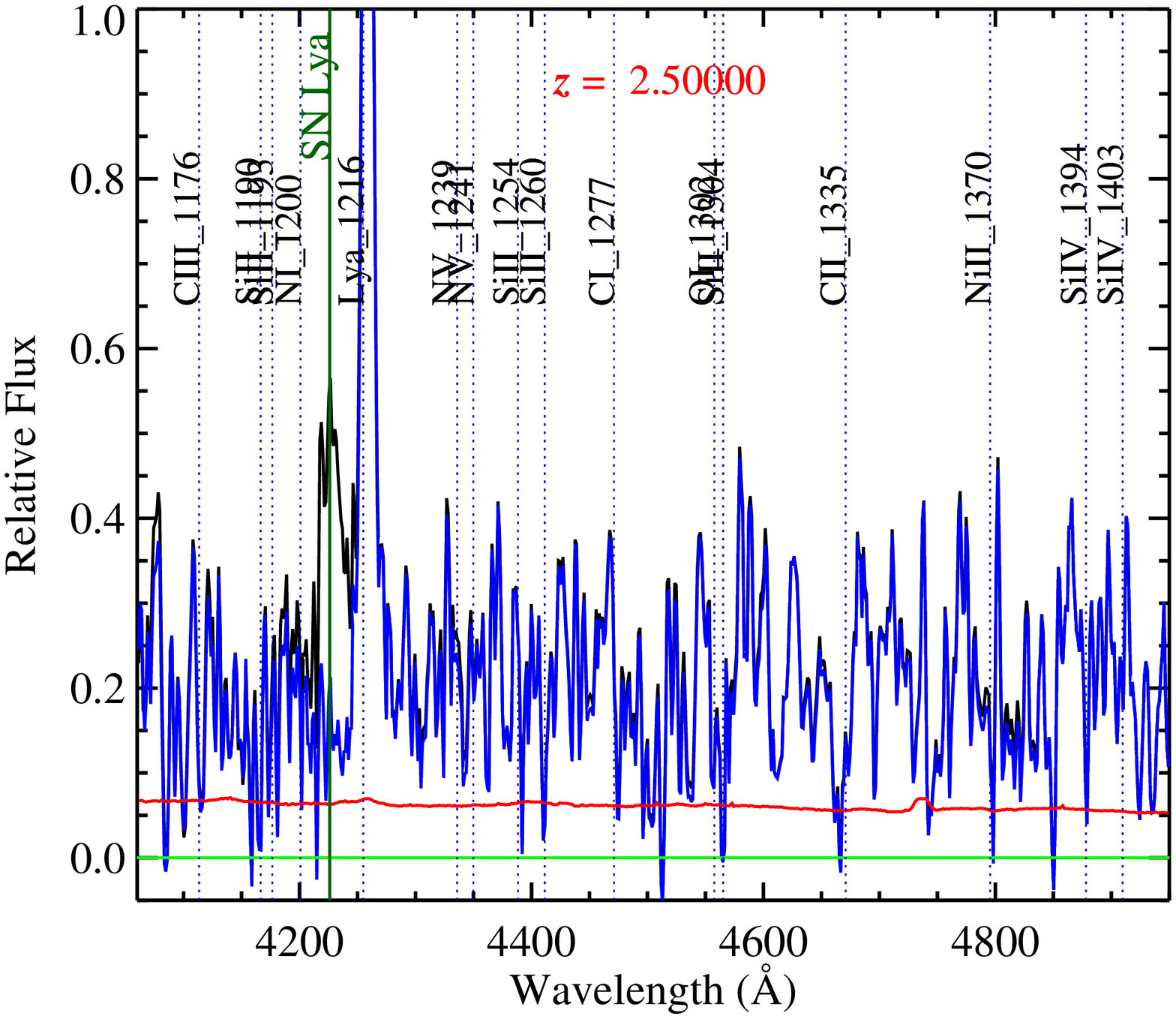}}}
\scalebox{0.5}[0.37]{\rotatebox{90}{\includegraphics{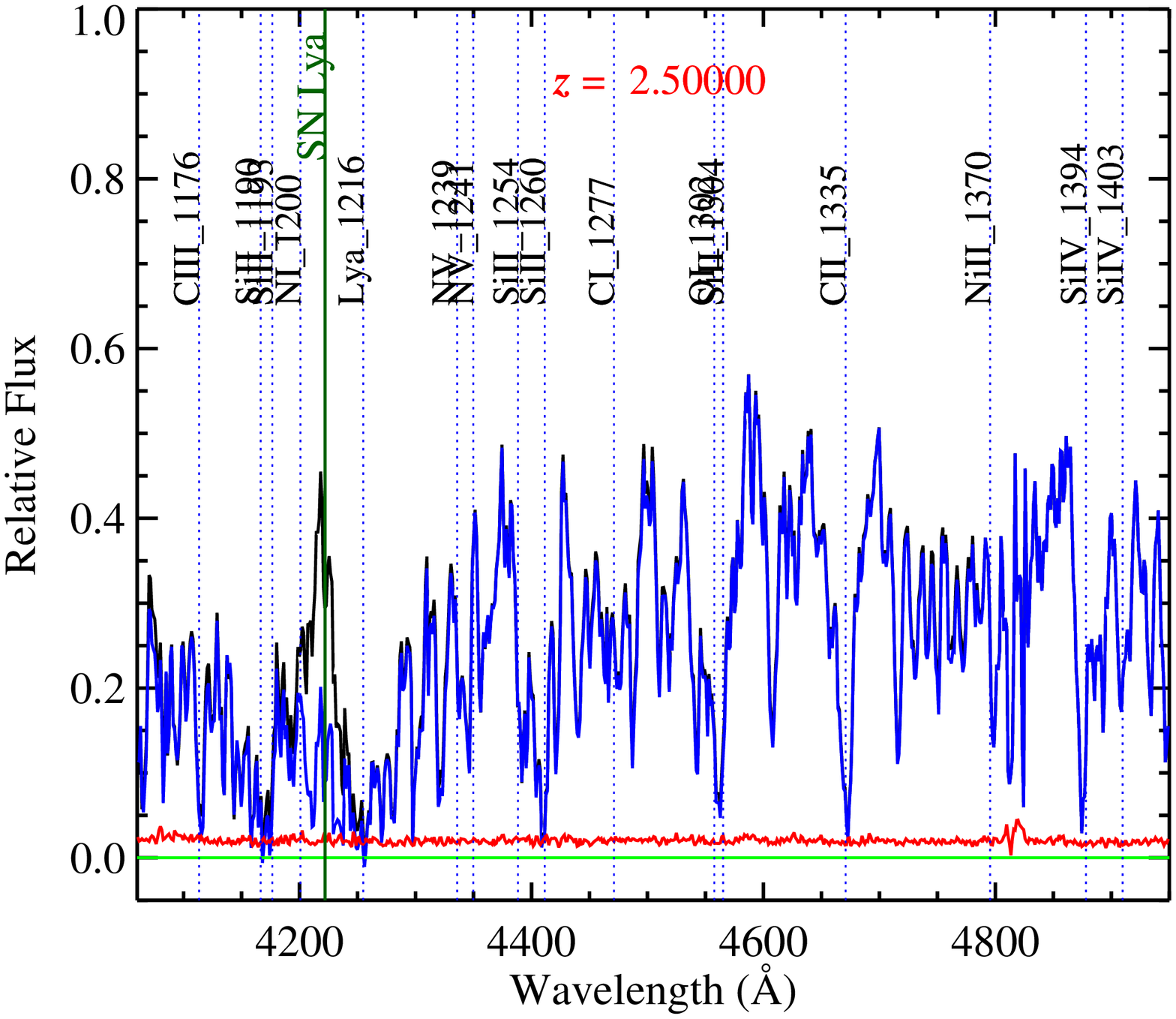}}}
%\scalebox{0.5}[0.38]{\rotatebox{90}{\includegraphics{f5a_BW.ps}}}
%\scalebox{0.5}[0.38]{\rotatebox{90}{\includegraphics{f5b_BW.ps}}}
\caption
{Spectra simulating $z=2.5$ Type IIn SNe detections.  The galaxy
  spectra are shown in black and the galaxy+SN spectra are shown in
  (blue) gray.  The spectra are estimates of peak to late Type IIn SNe
  Ly-$\alpha$ emission in simulated four hour integrations using
  existing 8m-class telescopes.  The LBG spectra displaying
  Ly-$\alpha$ in emission {\it (upper panel)} and Ly-$\alpha$ in
  absorption {\it (lower panel)} are based on actual data from
  \citet{c05}. Galaxy interstellar emission and absorption lines are
  labeled and marked by dotted vertical {\it (blue)} lines and the
  Ly-$\alpha$ emission lines from the Type IIn SNe are marked with
  solid vertical {\it (dark green)} lines. The SNe emission features
  are blueshifted by $\sim2000$ km s$^{-1}$ with respect to the
  systemic LBG redshift and illustrate the separation expected from
  their respective host galaxy Ly-$\alpha$ feature.  SNe are confirmed
  if their Ly-$\alpha$ emission evolves appropriately in subsequent
  observations.}
\label{fig:sim}
\end{center}
\end{figure}

The observed SNe IIn Ly-$\alpha$ blueshift is great enough to allow
the escape of photons from star forming regions and line-of-sight
interstellar clouds containing high column densities of neutral
hydrogen.  For example, a SN Ly-$\alpha$ feature blueshift of
$\sim$1000 km s$^{-1}$ is large enough for $\gtrsim90$\% of the
photons to escape systems with log N(H$\textsc{i}$) = 19 \citep{m07}
and a blueshift of $\sim$3000 km s$^{-1}$ can escape sightlines with
damped absorption [log N(H$\textsc{i}$) = 20.3].  From the surveys of
\citep{s03,c05}, it can be seen that LBGs exhibit a smooth range of
\lya profiles with $\sim$50\% dominated by \lya in emission and
$\sim$50\% dominated by \lya in absorption of varying EW.  From the
results of \citet{aes03} and an analysis of the spectra of \citet{c05}
it can be deduced that $>$90\% of the Ly-$\alpha$ photons of SNe IIn
can escape $\ge90$\% of all $1.9<z<3.3$ LBGs.  Therefore, assuming the
emission line evolution above is representative of the SNe IIn
population, we estimate that $\gtrsim$70\% of all $1.9<z<3.3$ SNe IIn
are spectroscopically detectable.  Figure~\ref{fig:sim} displays a
simulated detection of the peak \lya emission $(\sim-1500$ km
s$^{-1}$) from a SN IIn in two $z>2$ LBGs. The blueshifted SN \lya
emission feature is cleanly separated from the both the host $z=2.8$
LBG \lya emission feature and the host $3.1$ LBG \lya absorption
feature.

In summary, follow-up spectroscopic detection of SNe IIn Ly-$\alpha$
emission in $1.9<z<3.3$ LBGs is currently feasible and would identify
photometrically selected SNe candidates.  For \zrange SNe IIn, there
are no bright night sky emission lines to interfere with \lya emission
line detections. The blueshift of the SNe \lya emission line provides
a natural means to separate the SNe from the host galaxy \lya feature.
Subsequent observations to monitor the evolution of the SN \lya
emission would provide definitive confirmation of $z>2$ SNe without
photometric redshifts and template fitting.

\subsection{Contaminants}\label{contaminants}%Section3.2

For photometrically selected LBGs at $z>2$, AGN are the primary source
contaminant to mimic SNe IIn transient events.  Removing AGN from the
photometric results will depend directly on the design of the imaging
program.  Surveys using the LBG color-selection techniques described
here that have follow-up spectroscopic identifications
\citep{s03,s04,c05} find $\sim3$\% of the LBG candidates to
m$_{r}=25.5$ to have AGN activity, or $\sim70$ AGN deg$^{-2}$.  Guided
by the faint AGN luminosity function [e.g., \citep{sm07,w03}], we
estimate $\sim200$ AGN deg$^{-2}$ to m$_{r}=27.0$.  In practice, the
final density of AGN that appear like \zrange SNe IIn will be less
than this and will depend on AGN that (1) vary at the detectable level
during observation, (2) exhibit only one variation over multiple
epochs, and (3) follow SNe IIn FUV light curve expectations.
Nevertheless, many AGN are expected to contaminate the $z>2$
photometric SNe candidates.  All but the faintest of these can be
identified and eliminated with the deep follow-up spectroscopy that is
necessary to confirm SNe IIn candidates.

Using the magnitude depths in five filters ($u$*$griz$) and the five
epochs of the CFHTLS, we make a coarse estimate that $\sim30$ AGN
deg$^{-2}$ will pass the above constraints and contaminate the CFHTLS
data.  This is a few times the number of estimated SNe IIn
(observed-frame) that would be detected from \zrange
(\S~\ref{predictions}) and implies $\sim25-50$\% efficiency for the
CFHTLS Deep survey in follow-up deep spectroscopy.  Future surveys
with filters, epochs, and depths better tailored to this method and
refined information regarding the FUV evolution of SNe IIn, will
improve this efficiency.  For example, AGN typically exhibit stronger
C\textsc{iv}$\lambda1550$ emission relative to Mg\textsc{ii}
$\lambda2800$ emission.  Further evidence supporting the observed
reverse behavior of strong Mg\textsc{ii} $\lambda2800$ emission and
weak C\textsc{iv}$\lambda1550$ emission in SNe IIn \citep{f02,f05}
would substantiate an effective method of using narrow-band filters to
detect SNe IIn candidates while filtering out AGN.  This method would
invoke deep imaging with filters centered on
C\textsc{iv}$\lambda1550$, Mg\textsc{ii} $\lambda2800$, and an
off-band over multiple epochs and use the
C\textsc{iv}$\lambda1550$/Mg\textsc{ii} $\lambda2800$ line ratios to
discriminate between AGN and SNe IIn.

For completeness, the recently classified Type IIa SNe \citep{d04},
exhibit properties similar to SNe IIn and are very bright events.  To
date, only five objects have been placed in this category: SN 2006gy
\citep{ofek06}, SN 2005gj \citep{a06}, SN 2002ic \citep{h03}, SN 1999E
\citep{r03}, and SN 1997cy \citep{g00}.  The typical maximum
brightness and behavior of Type IIa SNe is uncertain because of the
small sample, but have been observed as bright as M$_V = -22.2$
\citep{ofek06}.  These rare events display bright emission lines from
circumstellar interaction and would contribute to the total number of
detected $z>2$ SNe events.  With careful design and more data, Type
IIa SNe detections may prove to be an excellent diagnostic of high
redshift binary C+O white dwarfs that are currently believed to be
their progenitors.

%*********************************************************************

\section{Observational Expectations}\label{expectations}% Section 4

\subsection{Type IIn supernovae rate estimation}\label{SNR}% Section 4.1

Although the Type Ia supernova rate (SNR) has been measured out to
$z\sim1.6$, the Type II SNR has not been measured beyond $z=1$
\citep{dahlen04,scan05,sul06,n06}.  To estimate the expected number of
in dedicated deep searches monitoring $1.9<z<3.3$ LBGs, we first
estimate the Type II SNR for $1.9<z<3.3$ LBGs and use the IIn/II
fraction to estimate the number of SNe IIn LBG$^{-1}$ yr$^{-1}$.
Finally, we use the color-selected \zrange LBG densities from deep
optical surveys to calculate the number of SNe IIn deg$^{-2}$
yr$^{-1}$.

The SFR and IMF of LBGs are two fundamental ingredients in determining
the \zrange Type II SNR.  We adopt an average LBG SFR of $40
M_{\odot}$ yr$^{-1}$ after considering (1) a median SFR of 44
$M_{\odot}$ yr$^{-1}$ for a sample of 74 LBGs with $\langle
z\rangle=2.97\pm0.27$ determined by spectral synthesis modeling of
their IR photometry \citep{aes01} and (2) an average of $\sim$36
$M_{\odot}$ yr$^{-1}$ from a sample of 798 LBGs with $\langle
z\rangle=2.96\pm0.29$ determined by their rest-frame
$\sim$1700\AA~flux \citep{aes03}.  These SFRs were determined using a
Salpeter IMF \citep{sal55} with a mass range of $0.1-125M_{\odot}$ and
is therefore the IMF we adopt here.  Assuming no evolution in the
average LBG SFR over $1.9<z<3.3$ and that all stars with $\ge8
M_{\odot}$ become core-collapse SNe, we find $\sim0.105$ Type II SNe
LBG$^{-1}$ yr$^{-1}$.  As a check, we compare this with the rate
derived by using SNRs at $z\sim1$ and observed trends to higher
redshift found in the literature and find it to be consistent.  The
Type Ia SNRs reported in \citet {sul06} and a median value from \citet
{scan05}, assuming a Type Ia/II SNR ratio of $\sim$3
\citep{dahlen04,madau98}, results in the expectations of 0.107 and
0.120 Type II SNe in \zrange LBG$^{-1}$ yr$^{-1}$, respectively.

The effect of dust is important in this estimation.  \citet{dahlen99}
show that the expected number of core-collapse SNe at high redshift is
relatively insensitive to dust.  They test different reddening values
and show that, in order to be consistent with the measured UV flux at
high redshift, higher dust requires a higher SNR.  As a result, the
number of detectable SNe is largely unaffected.  In addition, their
models assume all galaxies are established disk systems in which the
main impact is from edge-on systems.  From the existing HST images
[e.g, \citet{law07}], this does not appear to be the case for most
LBGs.  Therefore, we consider the overall effect of dust on the
expected number of core-collapse SNe to be small and within the
uncertainties of the estimate below.

There are varying estimates of the IIn/II SNe fraction ranging from
$2-5$\% \citep{cap97,dahlen04} to $\sim$12\% from a tabulation of
online magnitude-limited SN survey discoveries (e.g., CfA)\footnote
{http://cfa-www.harvard.edu/oir/Research/supernova/SNarchive.html}
that are biased toward the brightest SNe.  A possible physical
constraint comes from the fact that LBVs with initial mass consistent
with Type IIn SNe progenitors ($\gtrsim80 M_{\odot}$) make up
$\sim2.1$\% of the core-collapse progenitors using the above IMF.  We
adopt this latter value for our estimate and conclude that there are
$\sim0.0022$ SNe IIn LBG$^{-1}$ yr$^{-1}$ at \zrange.  The bulk of the
uncertainty in this estimate is in the SFR measurements and the SNe
IIn fraction.  For the IR sample of \citet{aes01} we used the median
SFR, however, the mean SFR is $102 M_{\odot}$.  The fraction of LBGs
with very high SFRs (some are fit with $\sim1000 M_{\odot}$ yr$^{-1}$)
will dominate the total number of expected SNe IIn.  In addition,
because observations constrain the SNe IIn fraction from being much
lower than our adopted value, we consider 2.1\% to be a conservative
and 3\% as a plausible upper limit.  Using these values in determining
the uncertainty, we estimate that there is one SNe IIn in
$\sim450^{+70}_{-325}$ \zrange LBGs.  Finally, we comment on an
additional factor that is not incorporated in this estimation.  Most
LBGs are best fit using young starburst models
\citep{sy98,aes01,pap01}.  Depending on the analysis, $\sim25$\% to
$>50$\% of the fits have timescales since last starburst ($t_{sf}$) of
$t_{sf}\lesssim10-35$ Myr.  Such timescales are in line with the ages
of massive stars and would result in these LBGs displaying a higher
number of SNe, implying a higher number of expected SNe IIn.

\subsection{Predictions}\label{predictions}% Section 4.2

Imaging surveys complete to $m_{r}=27$ using the photometric technique
of \citet{s03,s04} and can color select $\sim$14 LBGs arcmin$^{-2}$ at
$z=2.20\pm0.32$ and $\sim$9 LBGs arcmin$^{-2}$ at $z=2.96\pm0.26$
\citep{saw05}.  This results in $\sim$65000 LBGs deg$^{-2}$ at
$1.9<z<3.3$ after accounting for the photometric selection efficiency
of each population and color-selection overlap.  Our estimation above
implies that there exists roughly $\sim145^{+375}_{-20}$
($45^{+160}_{-6}$) SNe IIn deg$^{-2}$ yr$^{-1}$ rest-frame
(observed-frame) in color-selected $1.9<z<3.3$ LBGs to $m_{r}=27$.
Incorporating the observed \zrange LBG density with redshift
\citep{s03,s04,c05}, the LBG and SNe IIn detection functions
(\S~\ref{sims}), and the SNe IIn magnitude distribution
(\S~\ref{detection}), we find 3 (1), 25 (8), and 130 (40) detectable
SNe IIn deg$^{-2}$ yr$^{-1}$ rest-frame (observed-frame) for surveys
complete to m$_r = 26$, 27, and 28, respectively, to $ge4\sigma$ flux
uncertainty.  Therefore, probing one magnitude shallower or deeper
than the m$_r=27$ depth discussed throughout this work makes a
significant difference in the number of detectable SNe IIn deg$^{-2}$
yr$^{-1}$.  From earlier arguments (\S~\ref{lya}), we predict that
$\sim$70\% of these SNe IIn are detectable using existing facilities
for $\sim$2 years rest-frame..

We now estimate the observational expectations for the CFHTLS Deep
Synoptic Survey that consists of four square-degree pointings over
five yearly observing seasons (2003-2008).  Each of the seasonal
stacked images are complete to m$_{r'}\sim27$ and obtained over
approximately six month intervals (6-9 weeks rest-frame) per year with
tight constraints on the image quality to provide consistent data
(\S~\ref{sims}).  Accounting for the dilution of the observed
magnitude and multiple detections over an $\sim$50d rest-frame stacked
observation, we expect to find $\sim4^{+10}_{-1}$ SNe deg$^{-2}$
season$^{-1}$ in the observed frame.  Therefore, for the complete
four-degree, five-season CFHTLS Deep Synoptic Survey, we expect
$\sim60-280$ photometrically detectable $1.9<z<3.3$ SNe IIn with
$\sim40-200$ of these expected to have detectable Ly-$\alpha$ emission
lines.

\begin{figure}
\begin{center}
\scalebox{0.4}[0.4]{\rotatebox{90}{\includegraphics{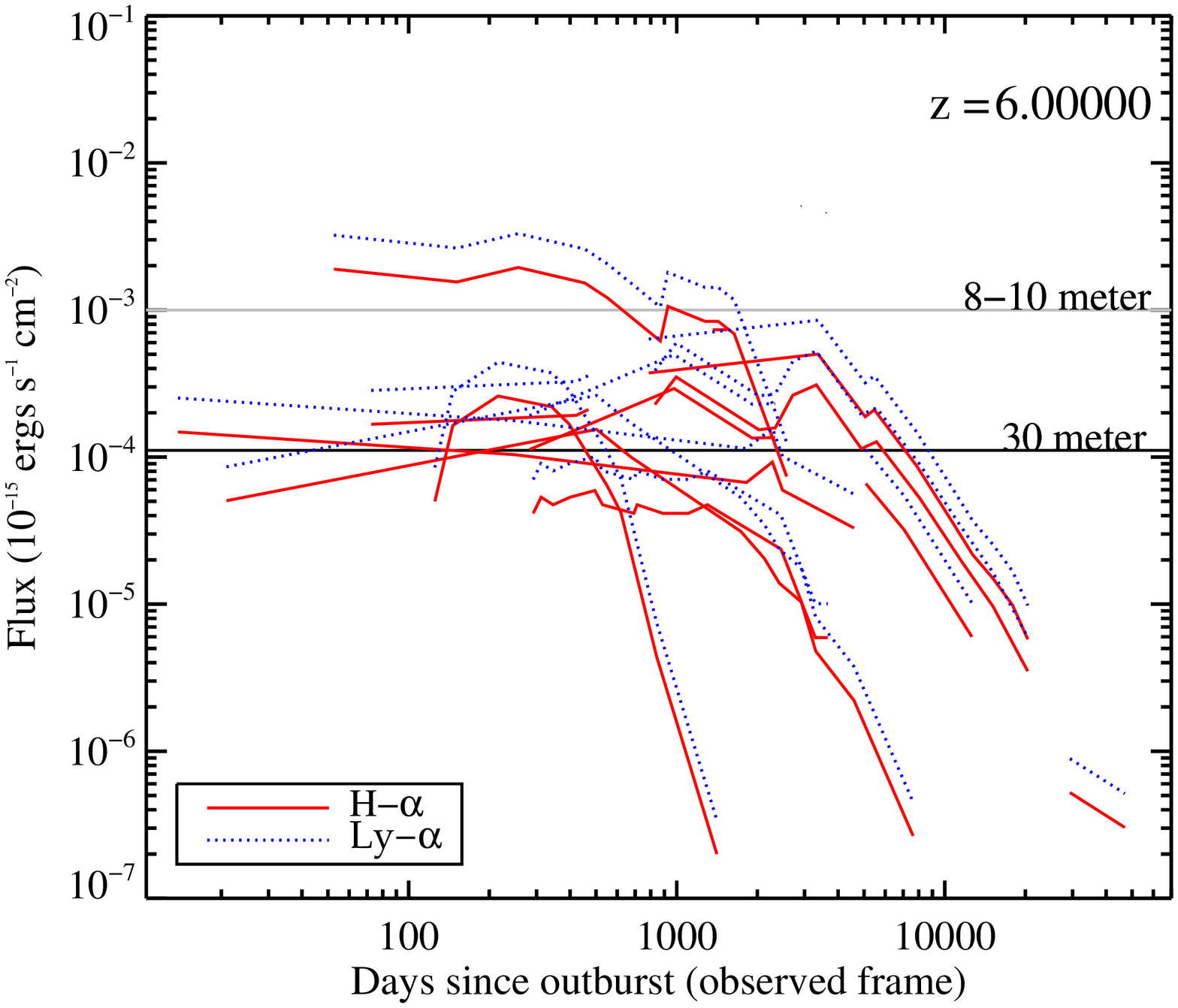}}}
%\scalebox{0.45}[0.45]{\rotatebox{90}{\includegraphics{f6_BW.ps}}}
\caption
{Same as Figure~\ref{fig:obj1250_z3} but with the H-$\alpha$ emission
  line flux evolution values redshifted to $z=6.0$. The sensitivity of
  future 30m-class telescopes have the potential to detect SNe IIn to
  $z\sim6$. }
\label{fig:z6}
\end{center}
\end{figure}

The capabilities of deep wide-field 8m-class telescope surveys and the
sensitivities of future 30m-class telescopes will detect a greater
density of \zrange SNe IIn.  It can be seen in
Figure~\ref{fig:obj1250_z3} that the projected sensitivities of future
facilities will spectroscopically confirm SNe emission at much greater
significance, for a longer duration, and will have the ability to
detect fainter SNe IIn emission lines.  Figure~\ref{fig:z6} displays
the results from low-redshift SNe IIn in Figure~\ref{fig:z01sn_flux}
corrected to $z=6$.  Inspection of this figure,
Figure~\ref{fig:obj1250_z3}, and the FUV spectra of SN1995N and
SN1998S \citep{f02,f05}, indicate that future facilities have the
potential to detect SNe IIn to $z\sim6$ as well as confirm $0<z<2$ SNe
IIn using Mg\textsc{ii}$\lambda2800$, H-$\alpha$, and other bright
lines to use SNe IIn to probe the universe from $0<z<6$.
 
%*********************************************************************

\section{Summary}\label{summary}% Section 5

We have presented a photometric and spectroscopic technique to detect
$z>2$ SNe IIn in reasonable numbers using current technology.
Although this technique can be applied to any redshift, we focus on
SNe at $1.9<z<3.3$ because of (1) the sensitivities and wide-field
capabilities of current optical instruments, (2) the success and
efficiency of $z>2$ photometric selection techniques, and (3) the
inherent properties of SNe IIn and $z>2$ galaxies.  

SNe IIn have $\langle M_B\rangle=19.0\pm0.9$ and very blue continua
with an estimate FUV (1700\AA) flux $\sim1$ mag fainter at maximum
brightness.  In addition, SNe IIn have bright emission features that
make a non-negligible contribution to the total flux that can be used
for detection and contaminant rejection.  We estimate that there exist
$\sim145^{+375}_{-20}$ SNe IIn deg$^{-2}$ yr$^{-1}$ rest-frame
($45^{+160}_{-6}$ observed-frame) in color-selected $1.9<z<3.3$ LBGs
to m$_r=27$.  We show from our simulations using real data obtained
from the CFHTLS Deep Synoptic Survey that deep, multi-band,
multi-epoch wide-field imaging surveys complete to m$_{r}\sim27$ can
detect $\sim$25 SNe IIn deg$^{-2}$ yr$^{-1}$ rest-frame ($\sim$8
observed-frame) at $1.9<z<3.3$ to $\ge4\sigma$ flux uncertainty.
These surveys can be performed independently or in conjunction with a
targeted narrow- to intermediate-band deep imaging program.
Contaminants, such as AGN, can be largely eliminated by the design of
the photometric program, selection criteria, and by follow-up deep
spectroscopy.

The general evolution of H-$\alpha$ and Ly-$\alpha$ emission lines at
low redshift predicts that the Ly-$\alpha$ features of $\gtrsim85$\%
of $z>2$ SNe IIn are above the spectroscopic thresholds of 8m-class
facilities for $\sim2$ years rest-frame.  Therefore, sensitive
low-resolution optical spectrographs on 8m-class telescopes have the
capability to spectroscopically identify $1.9<z<3.3$ SNe IIn
photometric candidates.  Positive confirmation of spectroscopic
candidates can be obtained at subsequent epochs as the line flux
evolves.  The blueshift ($\sim1500-2500$ km s$^{-1}$) of the \lya
emission lines observed in low redshift SNe IIn would result in a
clean separation from an LBG host galaxy Ly-$\alpha$ emission and
absorption if observed at $z>2$.  Under the assumption that this
property is inherent to SNe IIn, the findings of \citet{aes03} and the
analysis of the LBG spectroscopic sample of \citet{c05} result in an
estimate that the Ly-$\alpha$ emission lines from SNe IIn can escape
$\gtrsim$90\% of all $z>2$ LBGs.  Combined with the strength of SNe
IIn \lya emission, this results in the detectability of $\sim$70\% of
all $z>2$ SNe IIn.
 
To pursue the technique presented here, we have initiated a program to
better characterize SNe IIn features at low to intermediate redshift
using the {\it Hubble Space Telescope} and a complete program to
search for $1.9<z<3.3$ SNe IIn using the CFHTLS Deep Synoptic Survey
data.  Analyzing seasonal stacked images of the CFHTLS Deep Synoptic
survey, we estimate the photometric detection of $\sim4^{+6}_{-1}$ SNe
deg$^{-2}$ season$^{-1}$ in the observed frame, or $\sim60-200$ and
spectroscopic confirmation of $\sim40-140$ \zrange SNe IIn in the four
square-degree pointings over five years of the survey.  These, and
future programs, will pave the way for routine surveys using the next
generation of 8m-class survey and 30m-class telescopes that will
detect a greater density of SNe IIn and potentially extend SNe IIn
detection and Ly-$\alpha$ confirmation to $z\sim6$.  In addition, such
surveys could confirm $0<z<2$ SNe IIn using MgII, H-$\alpha$, and
other bright lines.  The detection of $z>2$ SNe IIn is important to
many areas of astrophysics, such as stellar physics and galaxy
formation and evolution, and will become more so as we learn more
about the mechanism behind SNe IIn and the details of their properties
and environments.

%*********************************************************************

\acknowledgments

The author would like to thank the referee for many helpful comments
and suggestions that greatly improved this paper and Mark Sullivan for
access to the deep images used in the simulations presented here and
helpful discussions.  The author would also like to thank Aaron Barth,
Betsy Barton, Douglas Leonard, Nao Suzuki, and Arthur Wolfe for
informative discussions and Kim Griest for the initial inspiration.
J.C. is supported in part by a Gary McCue Postdoctoral Fellowship
through the Center for Cosmology at the University of California,
Irvine.

%*********************************************************************

%*********************************************************************
%  Figures
%*********************************************************************

\clearpage

\clearpage

\clearpage

\clearpage

\clearpage

\clearpage


\begin{thebibliography}{}

\bibitem[Adelberger et al.(1998)]{a98}  Adelberger, K. L., Steidel,
  C. C., Giavalisco, M., Dickinson, M., Pettini, M., \& Kellogg,
  M. 1998, ApJ, 505, 18
\bibitem[Adelberger et al.(2003)]{a03} Adelberger, K. L., Steidel,
  C. C., Shapley, A. E., \& Pettini, M. 2003, ApJ, 584, 45
\bibitem[Aldering et al.(2006)]{a06} Aldering, G., Antilogus, P.,
  Bailey, S., Baltay, C., Bauer, A., Blanc, N., Bongard, S., Copin,
  Y., Gangler, E., Gilles, S., Kessler, R., Kocevski, D., Lee, B. C.,
  Loken, S., Nugent, P., Pain, R., P\'{e}contal, E., Pereira, R.,
  Perlmutter, S., Rabinowitz, D., Rigaudier, G., Scalzo, R., Smadja,
  G., Thomas, R. C., Wang, L. \& Weaver, B. A. 2006, ApJ, 650, 510
\bibitem[Ando(2004)]{ando04} Ando, S. 2004, ApJ, 607, 20
\bibitem[Appenzeller et al.(1998)]{ap98} Appenzeller, I., Fricke, K.,
  Furtig, W., Gassler, W., Hafner, R., Harkl, R., Hess, H.-J., Hummel,
  W., Jurgens, P., Kudritzki, R.-P., Mantel, K.-H., Meisl, W.,
  Muschielok, B., Nicklas, H., Rupprecht, G., Seifert, W., Stahl, O.,
  Szeifert, T. \& Tarantik, K. 1998, The Messenger, 94, 1
\bibitem[Aretxaga et al.(1999)]{a99} Aretxaga, I., Benetti, S.,
  Terlevich, R. J., Fabian, A. C., Cappellaro, E., Turatto, M. \&
  della Valle, M. 1999, MNRAS, 309, 343
\bibitem[Astier et al.(2006)]{astier06} Astier, P., Guy, J., Regnault,
  N., Pain, R., Aubourg, E., Balam, D., Basa, S., Carlberg, R. G.,
  Fabbro, S., Fouchez, D., Hook, I. M., Howell, D. A., Lafoux, H.,
  Neill, J. D., Palanque-Delabrouille, N., Perrett, K., Pritchet,
  C. J., Rich, J., Sullivan, M., Taillet, R., Aldering, G., Antilogus,
  P., Arsenijevic, V., Balland, C., Baumont, S., Bronder, J.,
  Courtois, H., Ellis, R. S., Filiol, M., Gonçalves, A. C., Goobar,
  A., Guide, D.,, Hardin, D., Lusset, V., Lidman, C., McMahon, R.,
  Mouchet, M., Mourao, A., Perlmutter, S., Ripoche, P., Tao, C., \&
  Walton, N. 2006, A\&A, 447, 31
\bibitem[Benetti et al.(1998)]{b98} Benetti, S., Cappellaro, E.,
  Danziger, I. J., Turatto, M., Patat, F., \& della Valle, M. 1998,
  MNRAS, 294, 448
\bibitem[Bertin \& Arnouts(1996)]{ba96} Bertin, E. \& Arnouts,
  S. 1996, A\&AS, 117, 393
\bibitem[Calura \& Matteucci(2006)]{calura06} Calura, F. \& Matteucci,
  F 2006, ApJ, 652, 889
\bibitem[Cappellaro et al.(1997)]{cap97} Cappellaro, E., Turatto, M.,
  Tsvetkov, D. Yu., Bartunov, O. S., Pollas, C., Evans, R. \& Hamuy,
  M. 1997, A\&A, 322, 431
\bibitem[Chugai et al.(1995)]{ch95} Chugai, N. N., Danziger, I. J. \&
  della Valle, M. 1995, MNRAS, 276, 530
\bibitem[Chugai (2001)]{chugai01} Chugai, N. N. 2001, MNRAS,
  326, 1448
\bibitem[Chugai et al.(2004)]{ch04} Chugai, N. N., Blinnikov, S. I.,
 Cumming, R. J., Lundqvist, P., Bragaglia, A., Filippenko, A. V.,
 Leonard, D. C., Matheson, T. \& Sollerman, J. 2004, MNRAS, 352, 1213
\bibitem[Cooke et al.(2005)]{c05} Cooke, J., Wolfe, A. M., Gawiser,
  E., \& Prochaska, J. X. 2005, ApJ, 621, 596
\bibitem[Cooke et al.(2006)]{c06} Cooke, J., Wolfe, A. M., Prochaska,
  J. X., \& Gawiser, E. 2006 ApJ, 652, 994
\bibitem[Dahlen \& Fransson(1999)]{dahlen99} Dahlen, T. \& Fransson,
  C. 1999 A\&A, 350, 349
\bibitem[Dahlen et al.(2004)]{dahlen04} Dahlen, T., Strolger, L. -G.,
  Riess, A. G., Mobasher, B., Chary, R. -R., Conselice, C. J.,
  Ferguson, H. C., Fruchter, A. S., Giavalisco, M., Livio, M., Madau,
  P., Panagia, N. \& Tonry, J. L. 2004, ApJ, 613, 189
\bibitem[Deng et al.(2004)]{d04} Deng, J., Kawabata, K. S., Ohyama,
  Y., Nomoto, K., Mazzali, P. A., Wang, L., Jeffery, D. J., Iye, M.,
  Tomita, H. \& Yoshii, Y. 2004, ApJ, 605, 37
\bibitem[Erb et al.(2004)]{erb04} Erb, D. K., Steidel, C. C., Shapley,
  A. E., Pettini, M., \& Adelberger, K. L. astro-ph/0404235
\bibitem[Fassia et al.(2001)]{f01} Fassia, A., Meikle, W. P. S.,
  Chugai, N., Geballe, T. R., Lundqvist, P., Walton, N. A., Pollacco,
  D., Veilleux, S., Wright, G. S., Pettini, M., Kerr, T.,
  Puchnarewicz, E., Puxley, P., Irwin, M., Packham, C., Smartt,
  S. J. \& Harmer, D. 2001, MNRAS, 325, 907
\bibitem[Filippenko(1997)]{f97} Filippenko, A. V. 1997, ARAA, 35, 309
\bibitem[Filippenko(1989)]{f89} Filippenko, A. V. 1989, AJ, 97, 726
\bibitem[Fransson et al.(2002)]{f02} Fransson, C., Chevalier, R. A.,
  Filippenko, A. V., Leibundgut, B., Barth, A. J., Fesen, R. A.,
  Kirshner, R. P., Leonard, D. C., Li, W., Lundqvist, P., Sollerman,
  J., Van Dyk \& S.  D. 2002, ApJ, 572, 350
\bibitem[Fransson et al.(2005)]{f05} Fransson, C., Challis, P. M.,
  Chevalier, R. A., Filippenko, A. V., Kirshner, R. P., Kozma, C.,
  Leonard, D. C., Matheson, T., Baron, E., Garnavich, P., Jha,
  S., Leibundgut, B., Lundqvist, P., Pun, C. S. J., Wang, L. \&
  Wheeler, J. C. 2005, ApJ, 622, 991
\bibitem[Fukugita et al.(1996)]{f96} Fukugita, M., Ichikawa, T., Gunn,
  J. E., Doi, M., Shimasaku, K., \& Schneider, D. P. 1996, AJ, 111,
  1748
\bibitem[Fukugita \& Kawasaki(2004)]{f04} Fukugita, M. \& Kawasaki,
  M. 2003, MNRAS, 340, 7
\bibitem[Gal-Yam et al.(2007)]{gal07} Gal-Yam, A., Leonard, D. C.,
  Fox, D. B., Cenko, S. B., Soderberg, A. M., Moon, D. -S., Sand,
  D. J., Li, W., Filippenko, A. V., Aldering, G. \& Copin, Y. 2007,
  656, 372
\bibitem[Gardner et al.(2001)]{g01} Gardner, J. P., Brown, T. M., \& 
 Ferguson, H. C. 2000, ApJ, 542L, 79
\bibitem[Germany et al.(2000)]{g00} Germany, L. M., Reiss, D. J.,
  Sadler, E. M., Schmidt, B, P. \& Stubbs, C. W. 2000, ApJ, 533, 320
\bibitem[Hagen et al.(1997)]{h97} Hagen, H.-J., Engels, D. \& Reimers,
  D. 1997, A\&A, 324, 29
\bibitem[Hamuy et al.(2003)]{h03} Hamuy, M., Phillips, M. M.,
  Suntzeff, N. B., Maza, J., Gonz\'{a}lez, L. E., Roth, M.,
  Krisciunas, K., Morrell, N., Green, E. M., Persson, S. E., \&
  McCarthy, P. J. 2003, Nature, 424, 651
\bibitem[Immler et al.(2006)]{i06} Immler, S. \& Pooley, D. 2006,
  ATel, 802, 1
\bibitem[Immler et al.(2007)]{i07} Immler, S. \& Brown, P. J. 2007,
  ATel, 1053, 1
\bibitem[Jung(2000)]{jung00} Jung, C. K. 2000, hep-ex/0005046  
\bibitem[Law et al.(2007)]{law07} Law, D. R., Steidel, C. C., Erb,
  D. K., Pettini, M., Reddy, N. A., Shapley, A. E., Adelberger, K. L.,
  \& Simenc, D. J. 2007, ApJ, 656, 1
\bibitem[Leonard et al.(2000)]{leonard00} Leonard, D. C., Filippenko,
  A. V., Barth, A. J., \& Matheson, T. 2000, ApJ, 536, 239
\bibitem[Li et al.(2002)]{li02} Li, W., Filippenko, A. V., van Dyk,
  S. D., Hu, J., Qiu, Y., Modiaz, M. \& Leonard, D. C. 2002,
  PASP, 114, 403
\bibitem[Li et al.(2007)]{li07} Li, W., Wang, X., Van Dyk, S. D.; 
  Cuillandre, J.-C., Foley, R. J. \& Filippenko, A. V. 2007, ApJ, 661,
  1013
\bibitem[Madau et al.(1998)]{madau98} Madau, P., Della Valle, M. \&
  Panagia, N. 1998, MNRAS, 297, L17 
\bibitem[Martin \& Sawicki(2004)]{martin04} Martin, C. L. \& Sawicki,
  M. 2004, ApJ, 603, 414
\bibitem[Mas-Hesse et al.(2007)]{m07} Mas-Hesse, J.M., Kunth, D.,
  Tenorio-Tagle, G., Leitherer, C., Terlevich, R. J. \& Terlevich,
  E. 2007 astro-ph/0309396
\bibitem[Nakamura(2003)]{nak03} Nakamura, K. 2003, J. Mod. Phys. A
  18, 4053
\bibitem[Neill et al.(2006)]{n06} Neill, J. D., Sullivan, M., Balam,
 D., Pritchet, C. J., Howell, D. A., Perrett, K., Astier, P., Aubourg,
 E., Basa, S., Carlberg, R. G., Conley, A., Fabbro, S., Fouchez, D.,
 Guy, J., Hook, I., Pain, R., Palanque-Delabrouille, N., Regnault, N.,
 Rich, J., Taillet, R., Aldering, G., Antilogus, P., Arsenijevic, V.,
 Balland, C., Baumont, S., Bronder, J., Ellis, R. S., Filiol, M.,
 Gon\c{c}alves, A. C., Hardin, D., Kowalski, M., Lidman, C., Lusset,
 V., Mouchet, M., Mourao, A., Perlmutter, S., Ripoche, P., Schlegel,
 D., \& Tao, C. 2006, AJ, 132, 1126
\bibitem[Ofek et al.(2006)]{ofek06} Ofek, E.O., Cameron, P. B.,
 Kasliwal, M. M., Gal-Yam, A., Rau, A., Kulkarni, S. R., Frail, D. A.,
 Chandra, P., Cenko, S. B., Soderberg, A. M., \& Immler, S. 2006,
 astro-ph/0612408
\bibitem[Oke et al.(1995)]{oke95} Oke, J. B., Cohen, J. G., Carr,
  M., Cromer, J., Dingizian, A., Harris, F. H., Labrecque, S.,
  Lucinio, R., Schaal, W., Epps, H., \& Miller, J. 1995, PASP, 107,
  375
\bibitem[Papovich et al.(2001)]{pap01} Papovich, C., Dickinson, M. \&
  Ferguson, Henry C. 2001, ApJ, 559, 620
\bibitem[Pastorello et al.(2002)]{p02} Pastorello, A., Turatto, M.,
  Benetti, S., Cappellaro, E., Danziger, I. J., Mazzali, P. A., Patat,
  F., Filippenko, A. V., Schlegel, D. J. \& Matheson, T. 2002, MNRAS,
  333 27
\bibitem[Pettini et a.(2001)]{pet01} Pettini, M., Shapley, A. E.,
 Steidel, C. C., Cuby, J-G., Dickinson, M., Moorwood, A. F. M.,
 Adelberger, K. L., \& Giavalisco, M. 2001, ApJ, 554, 981
\bibitem[Richardson et al.(2002)]{r02} Richardson, D., Branch, D.,
  Casebeer, D., Millard, J., Thomas, R. C., \& Baron, E. 2002, AJ,
  123, 745
\bibitem[Rigon et al.(2003)]{r03} Rigon, L., Turatto, M., Benetti, S.,
  Pastorello, A., Cappellaro, E., Aretxaga, I., Vega, O., Chavushyan,
  V., Patat, F., Danziger, I. J. \& Salvo, M. 2003, MNRAS, 340, 191
\bibitem[Salamanca et al.(2002)]{sal02} Salamanca, I., Terlevich,
  R. J. \& Tenorio-Tagle, G. S. 2002, MNRAS, 330, 844
\bibitem[Salpeter(1955)]{sal55} Salpeter, E. E. 1955, ApJ, 121, 161
\bibitem[Sawicki \& Thompson(2005)]{saw05} Sawicki, M., \& Thompson,
  D. 2005, ApJ, 635, 100
\bibitem[Sawicki \& Yee(1998)]{sy98} Sawicki, M., \& Yee,
  H. K. C. 1998, AJ, 115, 1329
\bibitem[Scannapieco \& Bildsten(2005)]{scan05} Scannapieco, E. \&
  Bildsten, L. 2005 ApJ, 629L, 85
\bibitem[Schlegel(1990)]{sch90} Schlegel, E. M. 1990, MNRAS, 244,
  269
\bibitem[Schlegel et al.(1999)]{sch99} Schlegel, E. M., Ryder,
  S., Staveley-Smith, L., Petre, R., Colbert, E., Dopita, M. \&
  Campbell-Wilson, D. 1999, AJ, 118, 2689
\bibitem[Shankar \& Mathur(2007)]{sm07} Shankar, F. \&  Mathur,
  S. 2007, ApJ, 660, 1051
\bibitem[Shapley et al.(2001)]{aes01} Shapley, A. E., Steidel,
  C. C., Adelberger, K. L., Dickinson, M., Giavalisco, M. \& Pettini,
  M. 2001, ApJ, 562, 95
\bibitem[Shapley et al.(2003)]{aes03} Shapley, A. E., Steidel,
  C. C., Pettini, M. \& Adelberger, K. L. 2003, ApJ, 588, 65
\bibitem[Steidel et al.(1996)]{s96} Steidel, C. C., Giavalisco, M.,
  Pettini, M., Dickinson, M., \& Adelberger, K. L. 1996, ApJ, 462L, 17
\bibitem[Steidel et al.(2003)]{s03} Steidel, C. C., Adelberger, K. L.,
  Shapley, A. E., Pettini, M., Dickinson, M., \& Giavalisco, M. 2003,
  ApJ, 592, 728
\bibitem[Steidel et al.(2004)]{s04} Steidel, C. C., Shapley,
  A. E., Pettini, M., Adelberger, K. L., Erb, D. K., Reddy,
  N. A., \& Hunt, M. P. 2004, ApJ, 604, 534
\bibitem[Strigari et al.(2004)]{louis04} Strigari, L. E., Kaplinghat,
  M., Steigman, G. \& Walker, T. P. 2004, JCAP, 03, 007
\bibitem[Sullivan et al.(2006)]{sul06} Sullivan, M., Le Borgne, D.,
  Pritchet, C. J., Hodsman, A., Neill, J. D., Howell, D. A., Carlberg,
  R. G., Astier, P., Aubourg, E., Balam, D., Basa, S., Conley, A.,
  Fabbro, S., Fouchez, D., Guy, J., Hook, I., Pain, R.,
  Palanque-Delabrouille, N., Perrett, K., Regnault, N., Rich, J.,
  Taillet, R., Baumont, S., Bronder, J., Ellis, R. S., Filiol, M.,
  Lusset, V., Perlmutter, S., Ripoche, P. \& Tao, C. 2006, ApJ, 648,
  868
\bibitem[Sullivan(2007)]{sul07} Sullivan, M., 2007, private
  communication.
\bibitem[Tapken et al.(2007)]{tap07} Tapken, C., Appenzeller, I.,
  Noll, S., Richling, S., Heidt, J., Meinkoehn, E. \& Mehlert,
  D. astro-ph/0702414 
\bibitem[Turatto et al.(1993)]{t93} Turatto, M., Cappellaro, E.,
  Danziger, I. J., Benetti, S., Gouiffes, C. \& della Valle, M. 1993,
  MNRAS 262, 128
\bibitem[Turatto et al.(2000)]{asiago00} Turatto, M., Suzuki, T.,
  Mazzali, P. A., Benetti, S., Cappellaro, E., Danziger, I. J.,
  Nomoto, K., Nakamura, T., Young, T. R., \& Patat, F. 2000, ApJ,
  534, 57  
\bibitem[also see van Dyk et al.(2006)]{vd06} Van Dyk, S. D., Li, W.,
  Filippenko, A. V., Humphreys, R. M., Chornock, R., Foley, R. \&
  Challis, P. M. 2006, astro-ph/0603025
\bibitem[Venemans, et al.(2007)]{v07} Venemans, B. P., R\"{o}ttgering,
  H. J. A., Miley, G. K., van Breugel, W. J. M., de Breuck, C., Kurk,
  J. D., Pentericci, L., Stanford, S. A., Overzier, R. A., Croft,
  S. \& Ford, H. 2007, A \& A, 461, 823
\bibitem[Wolf et al.(2003)]{w03} Wolf, C., Wisotzki, L., Borch, A.,
  Dye, S., Kleinheinrich, M., \& Meisenheimer, K. 2003, A\&A, 408, 499
\end{thebibliography}
\end{document}